\documentclass{article}

\usepackage[preprint]{neurips_2026}

\usepackage[utf8]{inputenc} 
\usepackage[T1]{fontenc}    
\usepackage{hyperref}       
\usepackage{url}            
\usepackage{booktabs}       
\usepackage{amsfonts}       
\usepackage{nicefrac}       
\usepackage{microtype}      
\usepackage{xcolor}         

\usepackage{amsmath}
\usepackage{amssymb}
\usepackage{mathtools}
\usepackage{amsthm}
\usepackage{subcaption}

\usepackage{bm}
\usepackage{stfloats}
\usepackage{caption}
\usepackage{tabularx}
\usepackage{xcolor}
\usepackage{tikz}
\usetikzlibrary{shapes.geometric, calc, fit, matrix, arrows.meta, arrows, positioning}
\usepackage{pgfplots}
\pgfplotsset{compat=1.18}
\usepgfplotslibrary{fillbetween}

\DeclareMathOperator{\clip}{clip}

\newcolumntype{C}{>{\centering\arraybackslash}X}
\newcolumntype{s}{>{\hsize=.3\hsize\linewidth=\hsize}C}
\newcolumntype{D}{>{\hsize=.4\hsize\linewidth=\hsize}C} 

\newcommand{\sizestd}{4}
\newcommand{\sizestdinter}{5}
\newcommand{\std}[1]{\fontsize{\sizestd}{\sizestdinter}\selectfont ±#1}
\newlength{\mycolspace}

\usepackage[most]{tcolorbox}
\usepackage[table]{xcolor}
\usepackage{amsthm, amsmath, enumitem}
\usepackage[capitalize,noabbrev]{cleveref}
\usepackage{wrapfig}
\usepackage{multirow}
\usepackage{placeins}
\definecolor{modcyan}{RGB}{200,245,245}

\newtcolorbox{modified}{
  colback=modcyan,
  colframe=modcyan,
  boxrule=0pt,
  arc=0pt,
  outer arc=0pt,
  left=4pt,
  right=4pt,
  top=4pt,
  bottom=4pt,
  breakable,
  float
}

\title{Decoupling Communication from Policy: Robust MARL under Bandwidth Constraints}

\author{%
  Alexi Canesse, Benoît Goupil, Jesse Read, Sonia Vanier \\
  École polytechnique (LIX), CNRS,\\
  Institut Polytechnique de Paris,\\
  Palaiseau, France\\
  \texttt{alexi.canesse@polytechnique.edu} \\
}

\begin{document}

\maketitle

\begin{abstract}
		Communication enables coordination in multi-agent reinforcement learning (MARL), but many real-world applications, e.g., search-and-rescue with drone swarms, operate under severe bandwidth constraints. 
		Many communication architectures still expose a coupled bottleneck in which a shared latent representation is used for both policy execution and inter-agent communication. Consequently, reducing message size directly limits the policy's latent space, often leading to significant performance degradation. 
		We address this with two contributions. First, we introduce \(\beta\), a normalised per-agent bandwidth budget that unifies sparsity, rounds, and message dimension into a single comparable constraint. 
		Second, we provide SLIM, a minimal architecture that decouples the communication pathway from the policy's latent representation, allowing us to isolate the effect of bandwidth from the effect of policy capacity while benefiting from in-step communication.
		We evaluate our method on several partially-observable MARL benchmarks, where communication is essential. Our approach achieves state-of-the-art performance and exhibits scalability and robustness under limited communication, with only marginal degradation as bandwidth is reduced.
\end{abstract}

\section{Introduction}

Multi-agent reinforcement learning (MARL) has seen significant advancements in recent years, enabling agents to learn cooperative behaviours in complex environments~\cite{sukhbaatar2016learning, das2019tarmac, yu2022surprising}. One of the main difficulties is effective cooperation between agents, especially in partially-observable settings where each agent has access to only limited information about the environment and other agents. Some methods rely on a central controller, effectively transforming the multi-agent problem into a single-agent one~\cite{claus1998dynamics, vinyals2019grandmaster}. However, such centralised approaches are often impractical in real-world applications due not only to scalability limitations (with the state and action spaces scaling with the number of agents) but also because such centralisation is simply not feasible; agents may be required to operate in a decentralised fashion. Centralised training with decentralised execution (CTDE)~\cite{lowe2017multi, kubatrust} has emerged as a popular paradigm, allowing agents to be trained with access to global information while executing policies based only on local observations.  Consequently, communication becomes essential to compensate for limited local observations.

\begin{figure}[t]
	\centering
	\resizebox{.8\textwidth}{!}{\input{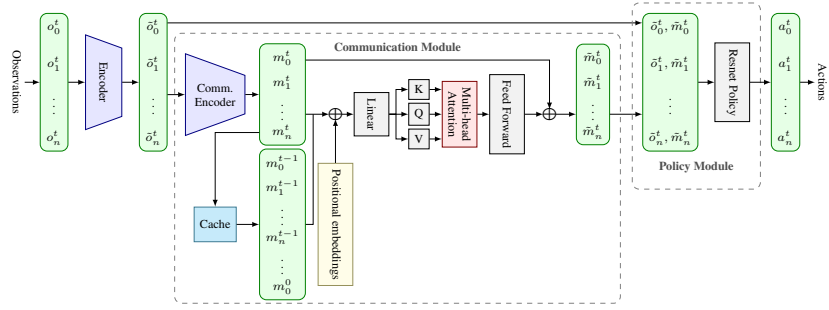}}
	\caption{\textbf{Architecture of the method.} Observations \((o^t_i)_{i,t}\) of agents \(\{i\}_{1, \dots, n}\) are encoded to reduce their dimension and increase their expressiveness before being passed to the \textrm{Communication Module} and the \textrm{Policy Module}. Crucially, encoded observations bypass the communication to reach the policy, allowing the communication module to reduce the dimension of the communication, through messages \((m_i^t)_{i,t}\), without loss of information for the policy.}%
	\label{fig:architecture}
\end{figure}

Communication is a mechanism that enables agents to get broader information about the environment and other agents. It enables agents to share information about their local observations, intentions, or learnt knowledge to improve cooperation and overall team performance. However, many real-world applications, such as search-and-rescue missions with drone swarms, autonomous vehicle fleets coordinating in traffic, or underwater exploration robots, involve severe bandwidth constraints. Reducing the size of communicated messages is essential for MARL systems to be deployed in such 
settings.

\begin{wrapfigure}[17]{r}{0.4\textwidth}
	\centering
	\resizebox{.31\textwidth}{!}{\begin{tikzpicture}[
    >=latex,
    vertex/.style={circle, draw, minimum size=.7cm, align=center, fill=blue!10!white, inner sep=1pt, font=\tiny},
    agent/.style={draw, minimum size=1.5cm, fill=blue!10!white},
    encoder/.style={draw=blue!20!black, fill=blue!50!red!15, minimum height=.1cm,
				shape=trapezium, trapezium angle=50, shape border rotate=270, align=center},
    ]
    
    \node[vertex] at (0,0) (agent1) {\(A_1\)};
    \node[vertex] at (1.5, 0) (agent2) {\(A_2\)};
    \node[vertex] at (0, -1.75) (agent3) {\(A_3\)};
    \node[vertex] at (1.5, -1.75) (agent4) {\(A_4\)};
    \node[vertex] at (2.5, -.875) (agent5) {\(A_5\)};
    \node[vertex] at (-1., -.875) (agent6) {\(A_6\)};

    \node[fit=(agent3)(agent4), draw, thick, inner sep=1em, fill=cyan!5!white, inner sep=1mm] (prezoomed) {};

    \node[vertex] at (0, -1.75) (agent3) {\(A_3\)};
    \node[vertex] at (1.5, -1.75) (agent4) {\(A_4\)};

    \draw[arrows={angle 60-angle 60}, thin] (agent1) -- (agent2);
    \draw[arrows={angle 60-angle 60}, , thin] (agent1) -- (agent3);
    \draw[arrows={angle 60-angle 60}, , thin] (agent2) -- (agent4);
    \draw[arrows={angle 60-angle 60}, , thin] (agent3) -- (agent4);
    \draw[arrows={angle 60-angle 60}, , thin] (agent1) -- (agent4);
    \draw[arrows={angle 60-angle 60}, , thin] (agent2) -- (agent3);  
    \draw[arrows={angle 60-angle 60}, , thin] (agent1) -- (agent5);
    \draw[arrows={angle 60-angle 60}, , thin] (agent2) -- (agent5);
    \draw[arrows={angle 60-angle 60}, , thin] (agent4) -- (agent5);
    \draw[arrows={angle 60-angle 60}, , thin] (agent1) -- (agent6);
    \draw[arrows={angle 60-angle 60}, , thin] (agent3) -- (agent6);
    \draw[arrows={angle 60-angle 60}, , thin] (agent6) -- (agent4);  
    \draw[arrows={angle 60-angle 60}, , thin] (agent6) -- (agent5);  
    \draw[arrows={angle 60-angle 60}, , thin] (agent6) -- (agent2);  
    \draw[arrows={angle 60-angle 60}, , thin] (agent5) -- (agent3);  

    \node[agent] at ($(agent3) - (-1.25cm, 2.25cm)$) (agent3b) {\(A_3\)};
    \node[agent, right=2cm of agent3b] (agent4b) {\(A_4\)};

    \node[encoder, anchor=center] at ($(agent3b.east) + (.5cm, .5cm)$) (encodera) {};
    \node[encoder, anchor=center, rotate=180] at ($(agent4b.west) - (.5cm, .5cm)$) (encoderb) {};

    \draw[->] ($(agent3b.east) + (0, .5cm)$) -- (encodera);
    \draw[->] ($(agent4b.west) - (0, .5cm)$) -- (encoderb);

    \node[inner sep=0mm] (obs3) at ($(agent3b.south) - (0, .5cm)$) {Obs.};
    \node[inner sep=0mm] (obs4) at ($(agent4b.south) - (0, .5cm)$) {Obs.};
    \node[inner sep=0mm] (act3) at ($(agent3b.north) + (0, .5cm)$) {Act.};
    \node[inner sep=0mm] (act4) at ($(agent4b.north) + (0, .5cm)$) {Act.};

    \draw[->] (obs3) -- (agent3b);
    \draw[->] (obs4) -- (agent4b);
    \draw[->] (agent3b) -- (act3);
    \draw[->] (agent4b) -- (act4);

    \node[fit=(agent3b)(agent4b)(obs3)(obs4)(act3)(act4), draw, thick, inner sep=2mm, fill=cyan!5!white] (zoomed) {};

    \draw[-, thick] ($(prezoomed.south west) + (0.015cm, 0.005cm)$) -- ($(zoomed.south west) - (-0.015cm, -0.004cm)$);
    \draw[-, thick] ($(prezoomed.north east) - (0.005cm, 0.013cm)$) -- ($(zoomed.north east) - (0.002cm, 0.01cm)$);

    \node[agent] at ($(agent3) - (-1.25cm, 2.25cm)$) (agent3b) {\(A_3\)};
    \node[agent, right=2cm of agent3b] (agent4b) {\(A_4\)};

    \node[encoder, anchor=center] at ($(agent3b.east) + (.5cm, .25cm)$) (encodera) {};
    \node[encoder, anchor=center, rotate=180] at ($(agent4b.west) - (.5cm, .25cm)$) (encoderb) {};

    \draw[double, double distance=12pt, green!50!black, fill=green!50!black] ($(agent3b.east) + (0, .25cm)$) -- (encodera);
    \draw[double, double distance=12pt, green!50!black, fill=green!50!black] ($(agent4b.west) - (0, .25cm)$) -- (encoderb);

    \draw[double, double distance=1pt, ->, red!90!black] (encoderb) -- ($(agent3b.east) - (0, .25cm)$) ;
    \draw[double, double distance=1pt, ->, red!90!black] (encodera) -- ($(agent4b.west) + (0, .25cm)$);

    \draw[->, green!50!black] ($(agent4b.west) - (0, .25cm)$) -- (encoderb);
    \draw[->, green!50!black] ($(agent3b.east) + (0, .25cm)$) -- (encodera);

    \node[inner sep=0mm] (obs3) at ($(agent3b.south) - (0, .5cm)$) {\footnotesize Observation};
    \node[inner sep=0mm] (obs4) at ($(agent4b.south) - (0, .5cm)$) {\footnotesize Observation};
    \node[inner sep=0mm] (act3) at ($(agent3b.north) + (0, .5cm)$) {\footnotesize Action};
    \node[inner sep=0mm] (act4) at ($(agent4b.north) + (0, .5cm)$) {\footnotesize Action};

    \draw[->] ($(obs3.north) - (0, .1mm)$) -- (agent3b);
    \draw[->] ($(obs4.north) - (0, .1mm)$) -- (agent4b);
    \draw[->] (agent3b) -- ($(act3.south) - (0, .5mm)$);
    \draw[->] (agent4b) -- ($(act4.south) - (0, .5mm)$);

    \node[] at ($(encodera)!.5!(encoderb) - (0, .8cm)$) {\({\color{red!90!black} d_\text{in}} < {\color{green!50!black} d_\text{out}}\)};

\end{tikzpicture}}
	\caption{\textbf{Problem setting.} Agents interact; receive local observations, exchange messages, and take actions to maximise rewards. We designed a method to reduce the size of the communicated messages with limited loss in performance.}%
    \label{fig:overview}
\end{wrapfigure}
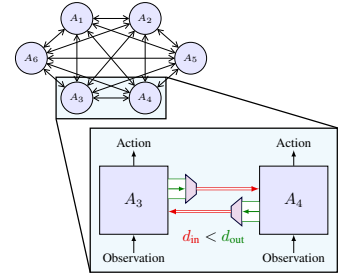

However, reducing communication without compromising on performance is a significant challenge. Machine learning models typically rely on high-dimensional feature spaces, and forcing them to operate in very low-dimensional spaces can lead to significant performance degradation~\cite{zhang2016understanding}. In MARL, many existing communication methods use a single shared latent representation both for the agent's policy and for the message it transmits~\cite{sukhbaatar2016learning, singh2018learning, hu2024learning}. In that regime, shrinking the message dimension imposes a double penalty: it removes information from transmitted messages and simultaneously reduces the representational capacity available to the policy itself, degrading overall performance. Some methods~\cite{das2019tarmac} already use a dedicated communication pathway but they communicate between timesteps, which introduces a lag in the communication. This issue is often solved using multi-round communication, but that is not always possible and increases the communication cost.

When deploying these systems in the real world, wireless networks impose two primary constraints: contention for medium access and bandwidth limitations~\cite{rappaport2010wireless}. The former requires sparsity in the communication graph (\textit{i.e.}, reducing the frequency or number of connections) while the latter can be addressed by employing either sparsity or smaller messages. The main focus of this paper is on the second constraint (as displayed in~\Cref{fig:overview}). We posit that optimising the size of individual messages is a more effective strategy for preserving coordination under bandwidth constraints.

To address this, we introduce Subdivided Lightweight Inter-agent Messaging (SLIM), a method that explicitly decouples communication from the policy input via a dedicated communication module. This separation allows the message dimension to be drastically reduced without restricting the policy's high-dimensional latent representation. We evaluate our approach across several partially observable MARL environments where communication is essential. Our method achieves state-of-the-art performance in high-bandwidth settings, while maintaining robust performance even as the message dimension is constrained to minimal values.

Concretely, we provide the following contributions:
\begin{itemize}

    \item{We establish a standardised evaluation protocol for communication-efficient MARL by defining a normalised bandwidth limit \(\beta\). By unifying message size, transmission rate, and graph sparsity into one constraint, we provide a framework for systematic benchmarking of communication strategies under identical physical bandwidth limitations.}

    \item{We propose a communication architecture tailored to within-timestep MARL communication that isolates message compression within a dedicated module, allowing message dimension to be reduced without constraining the higher-dimensional latent representation used for action selection.}
    
	\item {We provide an empirical study across four partially observable MARL benchmarks and several established baselines, showing that SLIM is consistently competitive at high bandwidth, notably more robust as bandwidth decreases, and that its cache is useful in environments that are not jointly fully observable.}%
\end{itemize}

\section{Related Work}

\paragraph*{Multi-agent Reinforcement Learning.} A straightforward approach to MARL treats each agent as an independent learner (IQL), applying standard reinforcement learning algorithms to each agent while treating others as part of the environment dynamics~\cite{tan1993multi, matignon2012independent}. This often leads to instability because the effective environment becomes non-stationary as other agents’ policies change. To address this, the centralised training with decentralised execution (CTDE) paradigm has emerged~\cite{lowe2017multi, foerster2018counterfactual}. It allows the training process to take advantage of global information, typically through a centralised critic or joint value function\cite{yu2022surprising} or value-factorisation methods~\cite{rashid2020monotonic}, while ensuring agents rely only on local observations during decentralised execution. Our training framework follows this paradigm, using a PPO ~\cite{schulman2017proximal} actor–critic objective with a centralised value function to compute advantages.

\paragraph*{Communication in MARL.} Many real-world applications of MARL involve partial observability, making inter-agent communication crucial for effective coordination and decision-making. Several works focus on discrete or interpretable communication~\cite{foerster2016learning, lazaridou2017multi, mordatch2018emergence, su2025goal} to address bandwidth limitations or accommodate low-power hardware. In contrast, our approach belongs to the category of continuous and differentiable communication. Within this domain, other methods have been proposed, such as different pooling techniques~\cite{sukhbaatar2016learning, singh2018learning} and graph neural networks~\cite{jiang2018graph, li2021deep, seraj2022learning, 10.1016/j.neucom.2024.127638, ding2023robust, jiang2024multi} to model agent interactions. Currently, attention mechanisms have emerged as the standard for MARL communication, whether integrated directly into the policy~\cite{hu2024learning} or implemented as a separate module~\cite{das2019tarmac}.

CommNet~\cite{sukhbaatar2016learning} introduced continuous communication in MARL by averaging agent hidden states to communicate between time steps. IC3Net~\cite{singh2018learning} builds upon this using individual rewards to mitigate credit assignment challenges~\cite{foerster2018counterfactual}. They also introduced a gating mechanism that enables the agents to learn when to communicate, further reducing communication throughput. However, in fully cooperative settings, this mechanism provides limited incentives to suppress unnecessary communication. Gated-ACML~\cite{mao2020learning} addresses this limitation by using Q-value differences to decide when to communicate, facilitating the pruning of uninformative messages.

Similarly to our approach, TarMAC~\cite{das2019tarmac} has a dedicated communication module separated from the policy's latent space. They also use an attention mechanism for agent interaction. However, their communication protocol is executed between time steps, which introduces a one-step temporal delay in information exchange. While this can be mitigated through multi-round communication, such a strategy incurs a linear increase in bandwidth consumption relative to the number of rounds.

In contrast to dense communication paradigms, Commformer~\cite{hu2024learning} introduces sparsity in the communication graph. Several works extend this by making the graph dynamic, allowing agents to selectively identify recipients or adapt content~\cite{jiang2018learning, ding2020learning, wang2021tom2c, wanglearning, rangwala2020learning, kubatrust,li2024context}. QLBT~\cite{9681886} goes further by taking into account network settings such as latency and saturation to adapt the communication graph between agents. Beyond topological adjustments, Sun et al.~\cite{sun2024dynamic} adapt message dimensionality to available bandwidth, while SchedNet~\cite{kim2019learning} frames agent selection as a scheduling problem, where only a subset of agents is permitted to transmit at each time step. Event-triggered methods~\cite{zhang2019efficient, hu2021event, han2023model} reduce transmission frequency by only communicating during critical state transitions. While effective, these approaches primarily focus on mitigating channel contention rather than reducing the dimensionality of individual messages, which is the central focus of our work. We consider these objectives to be complementary; combining temporal or topological sparsity with our proposed compression framework could further minimise overall communication overhead.

\paragraph*{Information Theory.} Bandwidth limitations in MARL communication have also been studied from information-theoretic perspectives~\cite{pmlr-v119-wang20i, ding2023robust, 10342771}. Such approaches typically aim to produce compact and informative messages, 
often motivated by classic results of transmission-rate constraints and principles of source coding ~\cite{shannon1948mathematical}. 
Methods from this literature operate at the level of message encoding (in contrast to our architectural considerations at the representation and policy level); while not directly within the scope of our work, they could potentially be integrated together in a multi-agent system, for greater communication efficiency.

\section{Subdivided Lightweight Inter-agent Messaging (SLIM)}

In this section we introduce our SLIM architecture. We first present the notation and preliminaries for MARL with communication. We then describe the SLIM architecture in detail, followed by the training procedure.

\subsection{Notation and Preliminaries}
\label{subsec:notations}

We consider multi-agent reinforcement learning (MARL) problems modelled as decentralised partially observable Markov decision processes (Dec-POMDPs)~\cite{oliehoek2016concise}, defined by the tuple \((\mathcal{S}, \mathcal{A}, P, R, Z, O, n, \gamma)\).
This tuple consists of a global state space \(\mathcal{S}\), a joint action space \(\mathcal{A} = \mathcal{A}_1 \times \ldots \times \mathcal{A}_n\) for \(n\) agents, state-transition dynamics \(P\), 
reward function \(R: \mathcal{S} \times \mathcal{A} \rightarrow \mathbb{R}^n\), and a discount factor \(\gamma \in [0, 1]\).
Partial observability of these processes is modeled by the joint observation space \(Z = Z_1 \times \ldots \times Z_n\) and the observation function \(O\). 

At discrete time \(t\), each agent \(i\) receives a local observation \(o_t^i \sim O( \cdot \mid s_{t}, a_{t-1})\)
(where $o_t^i \in Z_i$)
and selects an action \(a_t^i \sim \pi^i(\cdot \mid \tau_t^i)\) under its policy $\pi^i$ conditioned on its local action-observation history \(\tau_t^i = (o_0^i, a_0^i, \ldots, a_{t-1}^i, o_t^i)\).
The MARL objective is to learn a joint policy that maximises the expected cumulative discounted reward (return) \[
	J(\pi) = \mathbb{E}\left[\sum_{t=0}^{\infty} \gamma^t r_t\right].
\]

A Dec-POMDP is said to be \textbf{jointly fully observable} if the joint observation of all agents at all time \(t\), \(z_t = (o_t^1, \ldots, o_t^n)\), uniquely determines the true state \(s_t\) of the environment. A jointly fully observable Dec-POMDP is a decentralised Markov decision process (Dec-MDP)~\cite{oliehoek2016concise}.

Communication is allowed in these settings to mitigate partial observability. Some methods consider communication as part of the action space while others treat it as a separate mechanism. In order to generalise over different communication protocols, we decompose the agent's decision process into two distinct stages: message generation and transmission, followed by action selection. At each time step \(t\), each agent \(i\) first generates a message \(m_t^i\) based on its local history via a generation function \(\mu^i\).
This message is then transmitted to other agents. The agent then selects its action \(a_t^i\) conditioned on both its history and the set of messages received from the other agents.
\[
	m_t^i = \mu^i(\tau_t^i); \quad \quad a_t^i \sim \pi^i(\cdot \mid \tau_t^i, \{ m_t^j \}_{j \neq i}).
\]

This formulation unifies various communication protocols, where the received message set can represent direct signals, an aggregated mean field, or the output of a graph neural network. Furthermore, this process can be applied iteratively: steps 1 and 2 can be repeated \(k\) times within a single time step to allow for multi-round consensus or intent-based communication.

\subsection{The SLIM Architecture}

Our proposed SLIM architecture is illustrated in~\Cref{fig:architecture}. It implements the two-stage decision process using three main components: an observation encoder, a communication module, and a policy network. Non-jointly fully observable settings are specifically addressed through a message history cache integrated within the communication module.

\paragraph{Observation Encoding.}
The encoder \(E\) processes the local observation \(o_t^i\) of agent \(i\) at time step \(t\) to produce a latent representation \(\tilde{o}_t^i = E(o_t^i)\). This encoding step reduces the dimensionality of the observation while preserving relevant information for the policy.

\paragraph{Message Generation and Transmission.}
The representation \(\tilde{o}_t^i\) is passed to a communication encoder \(E_c\), which projects it into a compact message vector suitable for transmission defined by
\(
	m_t^i = E_c(\tilde{o}_t^i).
\)
These messages are then broadcasted to the other agents. Simultaneously, the agent receives the set of current messages \(\{ m_t^j \}_{j \neq i}\) from its peers.

\paragraph{Message History Cache.}
To address the partial observability of the environment, where the current observation alone is insufficient to determine the true state, each agent maintains a message cache \(\mathcal{C}_t^i\) potentially containing knowledge about the environment that could be useful later. This buffer stores the history of all exchanged messages in the system up to time \(t\):
\[
	\mathcal{C}_t^i = \mathcal{C}_{t-1}^i \cup \{ m_t^j \}_{1 \le j \le n} = \{m_{t'}^j\}_{1\leq j\leq n}^{1 \leq t' \leq t}.
\]
Because the cache stores only transmitted messages rather than full observation embeddings, its memory footprint scales linearly with the communication dimension \(d\).

\paragraph{Temporal Attention Aggregation.}
To synthesise the available information, we apply a transformer-based attention block over the full content of the cache \(\mathcal{C}_t^i\), which contains both the historical log and the messages just received at the current step. This mechanism dynamically weights the importance of each entry, enabling the agent to jointly reason over past and current information to construct a context vector \(\tilde{m}_t^i\).
Concretely, the transformer input is the sequence of cached messages augmented with two learnt positional embeddings: a temporal embedding that identifies the communication step and an agent embedding that identifies the sender. The current-step messages and the cached messages are processed by the same attention block, so \(\tilde{m}_t^i\) is an attention-weighted summary over who sent what and when, rather than a recurrent hidden state that privileges recent messages by construction.

\paragraph{Action Selection.}
Finally, the policy network \(\pi^i\) selects an action by conditioning on both the local representation and the context vector using
\[
	a_t^i \sim \pi^i(\tilde{o}_t^i, \tilde{m}_t^i).
\]

While the message history cache is critical for Dec-POMDPs, our architecture allows it to be optionally disabled. In Dec-MDP settings, where the current observation is sufficient for optimal decision-making, the cache mechanism can be deactivated. In this configuration, the attention block operates only on the current set of received messages.

\subsection{Centralised Training}

We follow the centralised training with decentralised execution paradigm. During training, we use a shared value function that receives the concatenated inputs from all agents. However, during execution, each agent only has access to its local observation and the messages received from other agents.

We use the standard MAPPO algorithm~\cite{yu2022surprising} to train our agents. Specifically, we apply the PPO objective~\cite{schulman2017proximal} to each agent individually irrespective of parameter sharing. For an agent \(i\) having policy parameters \(\theta_k\) with candidate policy parameters \(\theta\), an agent \(i\) with parameters \(\theta_k\) is updated to new parameters \(\theta\) by optimising the objective \(\mathcal L^p_i(\theta)\).
	{\begin{equation*}\begin{aligned}[t]
				\rho_{s, a}^{\theta_k} = \dfrac{\pi^i_\theta(a | s)}{\pi^i_{\theta_k}(a | s)}; \quad  \mathcal L^p_i(\theta) = & -\underset{s, a \sim \pi_\theta}{\mathbb{E}}\min \left(
				\rho_{s, a}^{\theta_k} A^i_{\pi_{\theta_k}}(s, a),\right.\left. \clip \left(\rho_{s, a}^{\theta_k}, 1-\varepsilon, 1+\varepsilon \right) A^i_{\pi_{\theta_k}}(s, a)\right)                                                             \\
				                                                                                                               & -\alpha\underset{s \sim \mathcal S}{\mathbb{E}} \sum_{a\in \mathcal A_i} \pi^i_\theta(a | s) \log \pi^i_\theta(a | s)
			\end{aligned}\end{equation*}
		where \(A^\cdot_{\pi_{\cdot}}(\cdot, \cdot)\) is the advantage function estimated using Generalised Advantage Estimation (GAE)~\cite{schulman2015high}, \(\varepsilon\) is a hyperparameter controlling the size of the policy update, \(\alpha\) is a temperature controlling the entropy bonus~\cite{haarnoja2018soft} that encourages exploration. A value estimation is necessary to compute the advantage estimates, and we train the value function \(V^i_\theta(\cdot)\) for each agent \(i\) by minimising the loss}
\[
	\mathcal L^v_i(\theta) = \underset{s, r \sim \pi_\theta}{\mathbb{E}} \left( V^i_\theta(s_t) - \hat R_t^i \right)^2.
\]
Since we are using centralised training, the state \(s_t\) used in the value function is the joint observation rather than the local observation of agent \(i\). To this end, our value function takes as input the concatenation of all agents' encoded observations and outputs a value estimate for each agent.

We can then combine those two losses to obtain the overall loss function per agent and take the mean over all agents to get the final loss function
\[
	\mathcal L(\theta, \theta_k) = \dfrac{1}{n}\sum_{i=1}^{n}\left(\mathcal L^p_i(\theta) + \mathcal L^v_i(\theta)\right).
\]

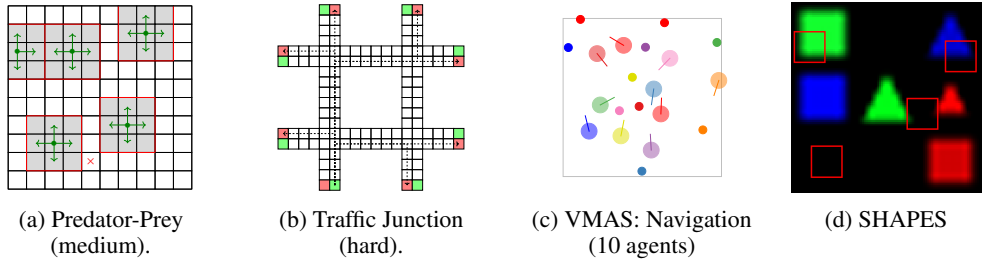
\begin{figure*}[t]
	\centering
	\hfill
	\begin{subfigure}[b]{0.225\textwidth}
		\centering
		\resizebox{!}{.8\linewidth}{\begin{tikzpicture}[scale=.4, transform shape]
	\matrix (pp) [matrix of nodes,
	nodes in empty cells,
	nodes={minimum width=2em, minimum height=2em, outer sep=0pt, anchor=south},
	column sep=-\pgflinewidth, row sep=-\pgflinewidth,
	cells={nodes={draw, font=\footnotesize}},
	] {
	& & & & & & |[fill=gray!30]| & |[fill=gray!30]| & |[fill=gray!30]| & \\
	|[fill=gray!30]| & |[fill=gray!30]| & |[fill=gray!30]| & |[fill=gray!30]| & |[fill=gray!30]| & & |[fill=gray!30]| & |[fill=gray!30]| \large\color{green!50!black}\(\bm{\bullet}\) & |[fill=gray!30]| & \\
	|[fill=gray!30]|\large\color{green!50!black}\(\bm{\bullet}\) & |[fill=gray!30]| & |[fill=gray!30]| & |[fill=gray!30]| \large\color{green!50!black}\(\bm{\bullet}\) & |[fill=gray!30]| & & |[fill=gray!30]| & |[fill=gray!30]| & |[fill=gray!30]| & \\
	|[fill=gray!30]| & |[fill=gray!30]| & |[fill=gray!30]| & |[fill=gray!30]| & |[fill=gray!30]| & & & & & \\
	& & & & & & & & & \\
	& & & & & |[fill=gray!30]| & |[fill=gray!30]| & |[fill=gray!30]| & & \\
	& |[fill=gray!30]| & |[fill=gray!30]| & |[fill=gray!30]| & & |[fill=gray!30]| & |[fill=gray!30]| \large\color{green!50!black}\(\bm{\bullet}\) & |[fill=gray!30]| & & \\
	& |[fill=gray!30]| & |[fill=gray!30]| \large\color{green!50!black}\(\bm{\bullet}\) & |[fill=gray!30]| & & |[fill=gray!30]| & |[fill=gray!30]| & |[fill=gray!30]| & & \\
	& |[fill=gray!30]| & |[fill=gray!30]| & |[fill=gray!30]| & \large\color{red}\(\bm{\times}\) & & & & & \\
	& & & & & & & & & \\
	};

	\draw[red, line width=1pt] (pp-1-1.south west) -- (pp-1-2.south east);
	\draw[red, line width=1pt] (pp-4-1.south west) -- (pp-4-2.south east);
	\draw[red, line width=1pt] (pp-4-3.south west) -- (pp-1-2.south east);

	\draw[red, line width=1pt] (pp-2-6.north west) -- (pp-4-5.south east);

	\draw[red, line width=1pt] (pp-2-3.north west) -- (pp-2-5.north east);
	\draw[red, line width=1pt] (pp-5-3.north west) -- (pp-5-5.north east);

	\draw[red, line width=1pt] (pp-1-7.north west) -- (pp-3-6.south east);
	\draw[red, line width=1pt] (pp-1-10.north west) -- (pp-3-9.south east);
	\draw[red, line width=1pt] (pp-4-7.north west) -- (pp-3-9.south east);

	\draw[red, line width=1pt] (pp-7-2.north west) -- (pp-6-4.south east);
	\draw[red, line width=1pt] (pp-7-2.north west) -- (pp-9-1.south east);
	\draw[red, line width=1pt] (pp-10-2.north west) -- (pp-9-4.south east);
	\draw[red, line width=1pt] (pp-7-5.north west) -- (pp-9-4.south east);

    \draw[red, line width=1pt] (pp-6-6.north west) -- (pp-5-8.south east);
	\draw[red, line width=1pt] (pp-6-6.north west) -- (pp-8-5.south east);
	\draw[red, line width=1pt] (pp-9-6.north west) -- (pp-8-8.south east);
	\draw[red, line width=1pt] (pp-6-9.north west) -- (pp-8-8.south east);

    \draw[black, line width=1pt] (pp-1-1.north west) -- (pp-1-10.north east);
	\draw[black, line width=1pt] (pp-1-1.north west) -- (pp-10-1.south west);
	\draw[black, line width=1pt] (pp-1-10.north east) -- (pp-10-10.south east);
	\draw[black, line width=1pt] (pp-10-1.south west) -- (pp-10-10.south east);

	\draw[->, green!50!black, line width=1pt] (pp-7-7.center) -- (pp-7-6.center);
	\draw[->, green!50!black, line width=1pt] (pp-7-7.center) -- (pp-8-7.center);
	\draw[->, green!50!black, line width=1pt] (pp-7-7.center) -- (pp-6-7.center);
	\draw[->, green!50!black, line width=1pt] (pp-7-7.center) -- (pp-7-8.center);

	\draw[->, green!50!black, line width=1pt] (pp-2-8.center) -- (pp-2-7.center);
	\draw[->, green!50!black, line width=1pt] (pp-2-8.center) -- (pp-3-8.center);
	\draw[->, green!50!black, line width=1pt] (pp-2-8.center) -- (pp-1-8.center);
	\draw[->, green!50!black, line width=1pt] (pp-2-8.center) -- (pp-2-9.center);

	\draw[->, green!50!black, line width=1pt] (pp-3-1.center) -- (pp-4-1.center);
	\draw[->, green!50!black, line width=1pt] (pp-3-1.center) -- (pp-2-1.center);
	\draw[->, green!50!black, line width=1pt] (pp-3-1.center) -- (pp-3-2.center);

	\draw[->, green!50!black, line width=1pt] (pp-3-4.center) -- (pp-3-3.center);
	\draw[->, green!50!black, line width=1pt] (pp-3-4.center) -- (pp-4-4.center);
	\draw[->, green!50!black, line width=1pt] (pp-3-4.center) -- (pp-2-4.center);
	\draw[->, green!50!black, line width=1pt] (pp-3-4.center) -- (pp-3-5.center);

	\draw[->, green!50!black, line width=1pt] (pp-8-3.center) -- (pp-8-2.center);
	\draw[->, green!50!black, line width=1pt] (pp-8-3.center) -- (pp-9-3.center);
	\draw[->, green!50!black, line width=1pt] (pp-8-3.center) -- (pp-7-3.center);
	\draw[->, green!50!black, line width=1pt] (pp-8-3.center) -- (pp-8-4.center);

\end{tikzpicture}} 
		\captionsetup{justification=centering}
		\caption{Predator-Prey\\ (medium).}\label{fig:pp}
	\end{subfigure}
	\hfill 
	\begin{subfigure}[b]{0.225\textwidth}
		\centering
		\resizebox{!}{.8\linewidth}{\begin{tikzpicture}
	\matrix (pp) [matrix of nodes,
	nodes in empty cells,
	nodes={minimum width=1.75em, minimum height=1.75em, outer sep=0pt, anchor=south},
	column sep=-\pgflinewidth, row sep=-\pgflinewidth,
	cells={nodes={font=\footnotesize}},
	] {
	\node {}; & \node {}; & \node {}; & \node {}; & \node[draw, fill=green!50] {}; & \node[draw, fill=red!50] (o00) {}; & \node {}; & \node {}; & \node {}; & \node {}; & \node {}; & \node {}; & \node[draw, fill=green!50] {}; & \node[draw, fill=red!50] (o01) {}; & \node {}; & \node {}; & \node {}; & \node {}; \\
	\node {}; & \node {}; & \node {}; & \node {}; & \node[draw] {}; & \node[draw] {}; & \node {}; & \node {}; & \node {}; & \node {}; & \node {}; & \node {}; & \node[draw] {}; & \node[draw] {}; & \node {}; & \node {}; & \node {}; & \node {}; \\
	\node {}; & \node {}; & \node {}; & \node {}; & \node[draw] {}; & \node[draw] {}; & \node {}; & \node {}; & \node {}; & \node {}; & \node {}; & \node {}; & \node[draw] {}; & \node[draw] {}; & \node {}; & \node {}; & \node {}; & \node {}; \\
	\node {}; & \node {}; & \node {}; & \node {}; & \node[draw] {}; & \node[draw] {}; & \node {}; & \node {}; & \node {}; & \node {}; & \node {}; & \node {}; & \node[draw] {}; & \node[draw] {}; & \node {}; & \node {}; & \node {}; & \node {}; \\
	\node[draw, fill=red!50] (o10) {}; & \node[draw] {}; & \node[draw] {}; & \node[draw] {}; & \node[draw] {}; & \node[draw] {}; & \node[draw] {}; & \node[draw] {}; & \node[draw] {}; & \node[draw] {}; & \node[draw] {}; & \node[draw] {}; & \node[draw] {}; & \node[draw] {}; & \node[draw] {}; & \node[draw] {}; & \node[draw] {}; & \node[draw, fill=green!50] {};\\
	\node[draw, fill=green!50] {}; & \node[draw] {}; & \node[draw] {}; & \node[draw] {}; & \node[draw] {}; & \node[draw] (m10) {}; & \node[draw] {}; & \node[draw] {}; & \node[draw] {}; & \node[draw] {}; & \node[draw] {}; & \node[draw] {}; & \node[draw] {}; & \node[draw] (m11) {}; & \node[draw] {}; & \node[draw] {}; & \node[draw] {}; & \node[draw, fill=red!50] (o11) {};\\
	\node {}; & \node {}; & \node {}; & \node {}; & \node[draw] {}; & \node[draw] {}; & \node {}; & \node {}; & \node {}; & \node {}; & \node {}; & \node {}; & \node[draw] {}; & \node[draw] {}; & \node {}; & \node {}; & \node {}; & \node {};\\
	\node {}; & \node {}; & \node {}; & \node {}; & \node[draw] {}; & \node[draw] {}; & \node {}; & \node {}; & \node {}; & \node {}; & \node {}; & \node {}; & \node[draw] {}; & \node[draw] {}; & \node {}; & \node {}; & \node {}; & \node {};\\
	\node {}; & \node {}; & \node {}; & \node {}; & \node[draw] {}; & \node[draw] {}; & \node {}; & \node {}; & \node {}; & \node {}; & \node {}; & \node {}; & \node[draw] {}; & \node[draw] {}; & \node {}; & \node {}; & \node {}; & \node {};\\
	\node {}; & \node {}; & \node {}; & \node {}; & \node[draw] {}; & \node[draw] {}; & \node {}; & \node {}; & \node {}; & \node {}; & \node {}; & \node {}; & \node[draw] {}; & \node[draw] {}; & \node {}; & \node {}; & \node {}; & \node {};\\
	\node {}; & \node {}; & \node {}; & \node {}; & \node[draw] {}; & \node[draw] {}; & \node {}; & \node {}; & \node {}; & \node {}; & \node {}; & \node {}; & \node[draw] {}; & \node[draw] {}; & \node {}; & \node {}; & \node {}; & \node {};\\
	\node {}; & \node {}; & \node {}; & \node {}; & \node[draw] {}; & \node[draw] {}; & \node {}; & \node {}; & \node {}; & \node {}; & \node {}; & \node {}; & \node[draw] {}; & \node[draw] {}; & \node {}; & \node {}; & \node {}; & \node {};\\
	\node[draw, fill=red!50] (o20) {}; & \node[draw] {}; & \node[draw] {}; & \node[draw] {}; & \node[draw] {}; & \node[draw] {}; & \node[draw] {}; & \node[draw] {}; & \node[draw] {}; & \node[draw] {}; & \node[draw] {}; & \node[draw] {}; & \node[draw] {}; & \node[draw] {}; & \node[draw] {}; & \node[draw] {}; & \node[draw] {}; & \node[draw, fill=green!50] {};\\
	\node[draw, fill=green!50] {}; & \node[draw] {}; & \node[draw] {}; & \node[draw] {}; & \node[draw] {}; & \node[draw] {}; & \node[draw] {}; & \node[draw] {}; & \node[draw] {}; & \node[draw] {}; & \node[draw] {}; & \node[draw] {}; & \node[draw] (m20) {}; & \node[draw] {}; & \node[draw] {}; & \node[draw] {}; & \node[draw] {}; & \node[draw, fill=red!50] (o21) {};\\
	\node {}; & \node {}; & \node {}; & \node {}; & \node[draw] {}; & \node[draw] {}; & \node {}; & \node {}; & \node {}; & \node {}; & \node {}; & \node {}; & \node[draw] {}; & \node[draw] {}; & \node {}; & \node {}; & \node {}; & \node {}; \\
	\node {}; & \node {}; & \node {}; & \node {}; & \node[draw] {}; & \node[draw] {}; & \node {}; & \node {}; & \node {}; & \node {}; & \node {}; & \node {}; & \node[draw] {}; & \node[draw] {}; & \node {}; & \node {}; & \node {}; & \node {}; \\
	\node {}; & \node {}; & \node {}; & \node {}; & \node[draw] {}; & \node[draw] {}; & \node {}; & \node {}; & \node {}; & \node {}; & \node {}; & \node {}; & \node[draw] {}; & \node[draw] {}; & \node {}; & \node {}; & \node {}; & \node {}; \\
	\node {}; & \node {}; & \node {}; & \node {}; & \node[draw, fill=red!50] {}; & \node[draw, fill=green!50] (i) {}; & \node {}; & \node {}; & \node {}; & \node {}; & \node {}; & \node {}; & \node[draw, fill=red!50] (o31) {}; & \node[draw, fill=green!50] {}; & \node {}; & \node {}; & \node {}; & \node {}; \\
	};

	\draw[->, dashed, line width=1pt] (i.center) -- (o00.center);
	\draw[->, dashed, line width=1pt] (i.center) |- (o10.center);
	\draw[->, dashed, line width=1pt] (i.center) |- (o11.center);
	\draw[->, dashed, line width=1pt] (i.center) |- (o20.center);
	\draw[->, dashed, line width=1pt] (i.center) |- (o21.center);
	\draw[->, dashed, line width=1pt] (i.center) -- (m10.center) -| (o01.center);
	\draw[->, dashed, line width=1pt] (i.center) |- (m20.center) -- (o31.center);
\end{tikzpicture}}
		\captionsetup{justification=centering}
		\caption{Traffic Junction\\ (hard).}\label{fig:tj}
	\end{subfigure}
	\hfill 
	\begin{subfigure}[b]{0.225\textwidth}
		\centering
		\resizebox{!}{.8\linewidth}{\includegraphics{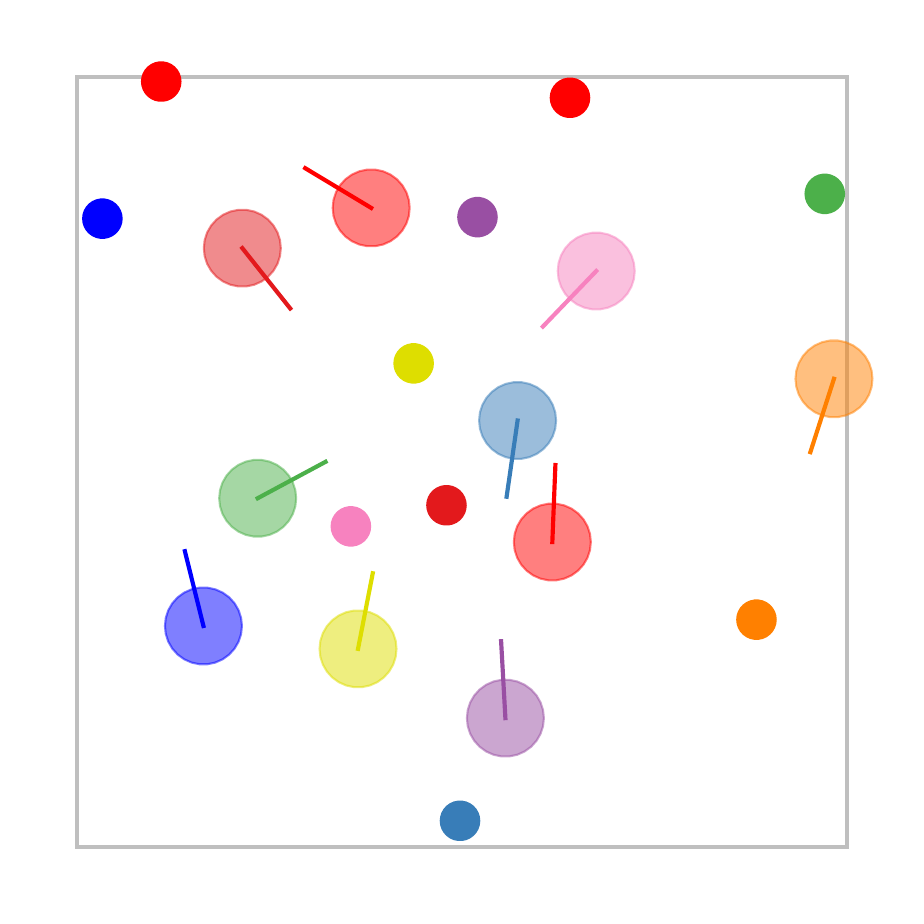}} 
        \captionsetup{justification=centering}
		\caption{VMAS: Navigation\\ (10 agents)}\label{fig:vmas}
	\end{subfigure}
	\begin{subfigure}[b]{0.225\textwidth}
		\centering
		\resizebox{!}{.8\linewidth}{\includegraphics{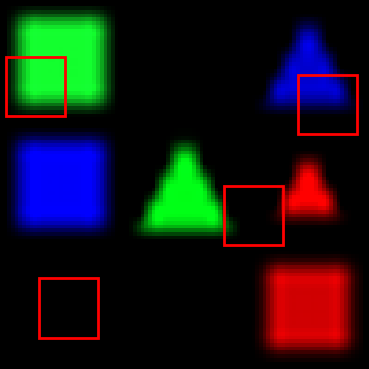}} 
        \captionsetup{justification=centering}
		\caption{SHAPES\\ \vphantom{text}}\label{img_shapes}
	\end{subfigure}
	\label{fig:three_envs}
	\caption{\textbf{Environments used in our experiments.} (\ref{fig:pp}) Predator-Prey: \(n\) predators cooperate to capture a fixed prey on a grid, each observing only a window (grey cells). (\ref{fig:tj}) Traffic Junction: cars navigate an intersection without vision, relying on communication to avoid collisions; green cells are spawn points, red cells goals, dashed arrows possible routes. Cars spawn with probability \(p\), up to a cap of \(n\). 
	(\ref{fig:vmas}) Navigation: \(n\) agents in a bounded continuous space must reach individually assigned goals (matching colour) without seeing peers. (\ref{img_shapes}) SHAPES: agents spawn on a random image of coloured shapes and must reach a target shape while observing only a local patch (red squares).}
	\label{fig:environments}
\end{figure*}

\section{Experiments}

In this section, we use multi-agent environments to evaluate the performance of our proposed method, the SLIM architecture, against several baselines. Our experiments aim to answer the following research questions: \(\bm{(i)}\) Under a high communication bandwidth $\beta$, does SLIM match or outperform representative baselines on multi-agent tasks? \(\bm{(ii)}\) How does SLIM’s performance evolve as the communication bandwidth budget is reduced over a broad range, and is it more robust than baseline methods in partially observable environments? \(\bm{(iii)}\) Does the cache help in improving performance in non-jointly observable environments?

\subsection{Normalised Agent Bandwidth}

We define the normalised agent bandwidth \(\beta\) as the maximum transmission capacity allocated to each agent (in floating-point scalars) per time step divided by the population. This value makes comparison easier between environments of varying number of agents. To ensure equivalent conditions, each model is subjected to the same normalised agent bandwidth. This constraint accounts for the message dimension \(d\), the message frequency \(k\) (corresponding to the number of communication rounds per time step as defined in~\Cref{subsec:notations}), and the sparsity of the communication graph \(\sigma\), defined as the fraction of the total population a given agent transmits to, into a single scalar limit. Formally, this bandwidth constraint is defined as
\begin{equation} \label{eq:bandwidth}
	\sigma \times k \times d \leq \beta.
\end{equation}
This standardisation allows us to compare different models at equal bandwidth limits, regardless of their specific communication strategy (e.g., trading sparsity in the communication graph for higher message resolution). Details on the computation of the bandwidth constraints for each model configuration are given in~\Cref{section:bandwidth_hyp}.

\subsection{Baselines}

We compare our SLIM architecture against a diverse set of established MARL baselines to cover a spectrum of communication strategies. We select CommNet~\cite{sukhbaatar2016learning} to represent a foundational continuous communication approach. We also include IC3Net~\cite{singh2018learning} which improves upon this architecture by using individual rewards and a gating mechanism.
However, consistent with observations in the original paper, we verified that in our fully cooperative scenarios, this gating mechanism remains effectively open, causing IC3Net to operate with a fully dense communication graph.
We further include TarMAC~\cite{das2019tarmac} because it is the closest prior architecture that already separates communication from the policy pathway.
Finally, we benchmark against CommFormer~\cite{hu2024learning}, a state-of-the-art transformer-based architecture. We evaluate both the dense variant (i.e., using a complete communication graph) to serve as a high-performance upper bound, and the sparse variant, which learns a fixed sparse communication graph, to assess the effectiveness of communication graph sparsification techniques as a bandwidth accommodation strategy. 

\subsection{Environments}


We evaluate on four standard partially-observable MARL benchmarks (illustrated in~\Cref{fig:environments}), with full descriptions in~\Cref{app:environments}. \textbf{Predator-Prey}~\cite{singh2018learning} is a grid world where predators with local vision cooperate to locate a stationary prey; since the prey does not move, past observations carry state information absent from the current joint observation, making the environment non-jointly-observable and motivating the message history cache (ablated in~\Cref{sec:ablation_cache}). \textbf{Traffic Junction}~\cite{singh2018learning} requires vision-less agents to cross an intersection along pre-assigned routes; concatenating all agents' positions fully determines the global state, so the environment is a Dec-MDP and the cache is disabled. \textbf{Navigation}~\cite{bettini2022vmas} places agents in a bounded continuous space where they must reach individual goals using acceleration actions under momentum, observing only their own position, velocity, and relative goal distance. \textbf{SHAPES}~\cite{andreas2016neural, das2019tarmac} spawns agents on an image of coloured shapes that must each reach a target colour while observing only a local patch; differing per-agent goals mean cached peer messages can carry goal-relevant information, providing a second testbed for the cache ablation.

\subsection{Experimental Setup}

\paragraph*{Reproducibility and Compute} All experiments are conducted across four seeds \(\{1, 2, 3, 4\}\) to ensure statistical robustness. We report mean performance alongside the standard error of the mean $(\text{std}/\sqrt{n_\text{seeds}})$ for all graphical representations and alongside the standard deviation for tabular results. The total computational budget for these experiments exceeded 5,000 GPU-hours on NVIDIA H100 GPUs. This is mostly due to the results necessary for~\Cref{sec:comm_efficiency}. While this number can seem high, it does not reflect inefficiencies, it is driven by the depth of our experiments, which required about 600 trainings (5 algorithm, 5 environments, 7 \(\beta\) values and 4 training seeds).

\paragraph*{Hyperparameter Protocol} For all baselines, we used the hyperparameters specified in their original publications after verifying their performance are unchanged at \(\beta=64\). For CommFormer, which was not originally evaluated on TrafficJunction, we performed a grid search over the PPO epoch parameter in \(\{5, 10, 15\}\) and the PPO clip value in \(\{0.05, 0.2\}\). These ranges were selected based on the values used in the original CommFormer experiments on other benchmarks. For SLIM, we performed a grid search, in high bandwidth, over the PPO epoch parameter in \(\{1, 5, 10\}\). The final values are provided in~\Cref{tab:slim_hyperparams} in the Appendix.

\subsection{Performance in High-Throughput Regimes}

We first establish a performance ceiling for all evaluated architectures by analysing them at the upper bandwidth bound ($\beta=2^6$). This regime represents a permissive setting where the communication channel is sufficiently wide, 
allowing us to assess the maximum expressive power of each model's communication protocol.
As detailed in \Cref{tab:high-band}, SLIM consistently achieves state-of-the-art performance across all benchmarks, either outperforming or maintaining statistical parity with established baselines. This verifies that SLIM's architectural design, while optimised for low-resource environments, does not sacrifice absolute performance when bandwidth is abundant.

\newlength{\mycolspacee}
\setlength{\mycolspacee}{2.5pt}
\begin{table}[t]
	\centering
	\caption{\textbf{Results on the different baselines using a bandwidth limit of \(\beta = 2^6\).} Performance on Predator-Prey is measured in average steps to capture the prey (lower is better), while performance on Traffic Junction is measured in success rate (higher is better) and in reward for Navigation (higher is better). (\(\dagger\)) Results for CommFormer are missing on the Navigation environment because it appears that the method is not able to converge.}
	\begin{tabularx}{\linewidth}{ccCCCCC}
		\toprule
		Environment                       & Difficulty & CommNet         & IC3Net          & TarMAC          & CommFormer              & \textbf{SLIM}             \\
		\midrule
		\multirow{2}{*}{Predator-Prey}    & Easy       & 5.68\std{0.48}  & 6.10\std{0.46}  & 6.18\std{0.50}  & \textbf{5.00\std{0.20}} & \textbf{4.97\std{0.04}}   \\
		                                  & Medium     & 24.78\std{0.99} & 20.14\std{2.52} & 17.86\std{3.46} & 14.06\std{1.54}         & \textbf{12.57\std{0.15}}  \\
		\midrule
		\multirow{2}{*}{Traffic Junction} & Easy       & 31.2\std{19.2}  & 85.8\std{10.0}  & 65.5\std{29.6}  & 84.5\std{13.2}          & \textbf{99.3\std{0.30}}   \\
		                                  & Medium     & 77.0\std{5.34}  & 80.6\std{7.42}  & 71.0\std{2.87}  & \textbf{96.0\std{4.31}} & \textbf{97.2\std{0.84}}   \\
		\midrule
		Navigation                        &            & 0.49\std{0.06}  & 0.28\std{0.24}  & 0.64\std{0.24}  & \(\dagger\)             & \textbf{0.81\std{0.05}} \\
		\bottomrule
	\end{tabularx}
	\label{tab:high-band}
\end{table}

\subsection{Robustness Under Various Bandwidth Constraints}\label{sec:comm_efficiency}

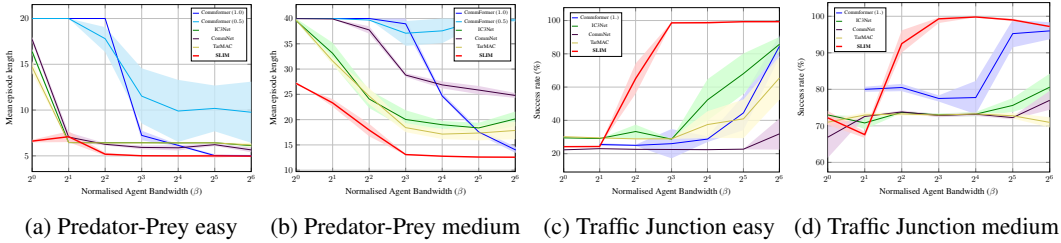
\begin{figure*} 
	\centering
	\begin{subfigure}[b]{0.244\textwidth}
		\centering
		\resizebox{\textwidth}{!}{\begin{tikzpicture}
	\begin{axis}[
			xlabel={Normalised Agent Bandwidth (\(\beta\))},
			ylabel={Mean episode length},
			xmode=log,
			xmin=1,
			xmax=64,
			log basis x=2,
			legend image post style={scale=0.4},
			legend style={
					font=\tiny,
					cells={align=center},
					inner sep=1pt
				},
			width=3.\textwidth,
			height=2.4\textwidth,
			grid=both,
			minor x tick num=1,
			xminorgrids=true,
		]
		\addplot[name path=upper, blue!80!black, forget plot, very thin, draw=none] coordinates {
				(2, 20)
				(4, 20)
				(8, 7.75)
				(16, 6.245)
				(32, 5.17)
				(64, 5.095)
			};
		\addplot[name path=lower, blue!80!black, forget plot, very thin, draw=none] coordinates {
				(2, 20)
				(4, 20)
				(8, 6.695)
				(16, 6.035)
				(32, 4.918)
				(64, 4.905)
			};

		\addplot[fill=blue!20, forget plot, opacity=0.8] fill between[of=upper and lower];

		\addplot[thick, blue] coordinates {
				(2, 20)
				(4, 20)
				(8, 7.235)
				(16, 6.1425)
				(32, 5.03)
				(64, 5.0)
			};
		\addlegendentry{Commformer (\(1.0\))}

		\addplot[name path=upper, cyan!80!black, forget plot, draw=none] coordinates {
				(1, 20)
				(2, 20)
				(4, 19.055)
				(8, 14.56)
				(16, 13.295)
				(32, 12.68)
				(64, 13.1)
			};
		\addplot[name path=lower, cyan!80!black, forget plot, very thin, draw=none] coordinates {
				(1, 20)
				(2, 20)
				(4, 16.335)
				(8, 8.54)
				(16, 6.525)
				(32, 7.68)
				(64, 6.42)
			};

		\addplot[fill=cyan!20, forget plot, opacity=0.8] fill between[of=upper and lower];

		\addplot[thick, cyan] coordinates {
				(1, 20)
				(2, 20)
				(4, 17.795)
				(8, 11.5225)
				(16, 9.895)
				(32, 10.1825)
				(64, 9.76)
			};
		\addlegendentry{Commformer (\(0.5\))}

		\addplot[name path=upper, green!80!black, forget plot, very thin, draw=none] coordinates {
				(1, 16.71)
				(2, 6.50)
				(4, 6.453)
				(8, 6.44)
				(16, 6.43)
				(32, 6.44)
				(64, 6.346)
			};
		\addplot[name path=lower, green!80!black, forget plot, very thin, draw=none] coordinates {
				(1, 16.06)
				(2, 6.39)
				(4, 6.418)
				(8, 6.42)
				(16, 6.39)
				(32, 6.41)
				(64, 5.88)
			};

		\addplot[fill=green!20, forget plot, opacity=0.8] fill between[of=upper and lower];

		\addplot[thick, green!50!black] coordinates {
				(1, 16.39)
				(2, 6.45)
				(4, 6.435)
				(8, 6.438)
				(16, 6.415)
				(32, 6.43)
				(64, 6.117)
			};
		\addlegendentry{IC3Net}

		\addplot[name path=upper, violet!80!black, forget plot, very thin, draw=none] coordinates {
				(1, 18.017)
				(2, 7.56)
				(4, 6.38)
				(8, 6.12)
				(16, 6.17)
				(32, 6.37)
				(64, 5.905)
			};
		\addplot[name path=lower, violet!80!black, forget plot, very thin, draw=none] coordinates {
				(1, 17.427)
				(2, 6.54)
				(4, 6.14)
				(8, 5.72)
				(16, 5.59)
				(32, 6.07)
				(64, 5.395)
			};

		\addplot[fill=violet!20, forget plot, opacity=0.8] fill between[of=upper and lower];

		\addplot[thick, violet!50!black] coordinates {
				(1, 17.722)
				(2, 7.05)
				(4, 6.26)
				(8, 5.92)
				(16, 5.88)
				(32, 6.22)
				(64, 5.65)
			};
		\addlegendentry{CommNet}

		\addplot[name path=upper, yellow!80!black, forget plot, very thin, draw=none] coordinates {
				(1, 15.365)
				(2, 6.47288)
				(4, 6.43823)
				(8, 6.42978)
				(16, 6.42302)
				(32, 6.43176)
				(64, 6.42741)
			};
		\addplot[name path=lower, yellow!80!black, forget plot, very thin, draw=none] coordinates {
				(1, 14.035)
				(2, 6.43288)
				(4, 6.38223)
				(8, 6.40848)
				(16, 6.40722)
				(32, 6.40516)
				(64, 5.93746)
			};

		\addplot[fill=yellow!20, forget plot, opacity=0.8] fill between[of=upper and lower];

		\addplot[thick, yellow!75!black] coordinates {
				(1, 14.70)
				(2, 6.45288)
				(4, 6.41023)
				(8, 6.41463)
				(16, 6.41512)
				(32, 6.41846)
				(64, 6.18241)
			};
		\addlegendentry{TarMAC}

		\addplot[name path=upper, red!80!black, forget plot, very thin, draw=none] coordinates {
				(1, 6.7536)
				(2, 7.675)
				(4, 5.35)
				(8, 5.0245)
				(16, 5.01)
				(32, 5.192)
				(64, 4.988)
			};
		\addplot[name path=lower, red!80!black, forget plot, very thin, draw=none] coordinates {
				(1, 6.4665)
				(2, 6.505)
				(4, 4.9975)
				(8, 5.05)
				(16, 4.975)
				(32, 4.802)
				(64, 4.952)
			};

		\addplot[fill=red!20, forget plot, opacity=0.8] fill between[of=upper and lower];

		\addplot[ultra thick, red] coordinates {
				(1, 6.61)
				(2, 7.09)
				(4, 5.18)
				(8, 5.011)
				(16, 4.995)
				(32, 4.997)
				(64, 4.97)
			};
		\addlegendentry{\textbf{SLIM}}

	\end{axis}
\end{tikzpicture}}
		\captionsetup{justification=centering}
		\caption{Predator-Prey easy}\label{fig:pp_easy_comm_eff}
	\end{subfigure}
	\begin{subfigure}[b]{0.244\textwidth}
		\centering
		\resizebox{\textwidth}{!}{\begin{tikzpicture}
	\begin{axis}[
			xlabel={Normalised Agent Bandwidth (\(\beta\))},
			ylabel={Mean episode length},
			xmin=1,
			xmax=64,
			xmode=log,
			log basis x=2,
			legend image post style={scale=0.4},
			legend style={
					font=\tiny,
					cells={align=center},
					inner sep=1pt
				},
			width=3.\textwidth,
			height=2.4\textwidth,
			grid=both,
			minor x tick num=1,
			xminorgrids=true,
		]
		\addplot[name path=upper, blue!80!black, forget plot, very thin, draw=none] coordinates {
				(2, 40)
				(4, 40)
				(8, 40)
				(16, 25.265)
				(32, 18.085)
				(64, 14.64)
				(128, 13.9)
			};
		\addplot[name path=lower, blue!80!black, forget plot, very thin, draw=none] coordinates {
				(2, 40)
				(4, 40)
				(8, 37.795)
				(16, 24.415)
				(32, 17.035)
				(64, 13.44)
				(128, 13.33)
			};

		\addplot[fill=blue!20, forget plot, opacity=0.8] fill between[of=upper and lower];

		\addplot[thick, blue] coordinates {
				(2, 40)
				(4, 40)
				(8, 38.945)
				(16, 24.68)
				(32, 17.56)
				(64, 14.04)
				(128, 13.33)
			};
		\addlegendentry{CommFormer (\(1.0\))};

		\addplot[name path=upper, cyan!80!black, forget plot, draw=none] coordinates {
				(1, 40)
				(2, 40)
				(4, 40)
				(8, 39.5)
				(16, 40)
				(32, 40)
				(64, 40)
			};
		\addplot[name path=lower, cyan!80!black, forget plot, very thin, draw=none] coordinates {
				(1, 40)
				(2, 40)
				(4, 39.84)
				(8, 34.6)
				(16, 35.12)
				(32, 38.88)
				(64, 39.365)
			};

		\addplot[fill=cyan!20, forget plot, opacity=0.8] fill between[of=upper and lower];

		\addplot[thick, cyan] coordinates {
				(1, 40)
				(2, 40)
				(4, 39.69)
				(8, 37.09)
				(16, 37.56)
				(32, 39.60)
				(64, 39.715)
			};
		\addlegendentry{CommFormer (\(0.5\))}

		\addplot[name path=upper, green!80!black, forget plot, very thin, draw=none] coordinates {
				(1, 40)
				(2, 35.21)
				(4, 25.85)
				(8, 21.85)
				(16, 19.735)
				(32, 19.385)
				(64, 21.597)
			};
		\addplot[name path=lower, green!80!black, forget plot, very thin, draw=none] coordinates {
				(1, 39.46)
				(2, 31.01)
				(4, 22.30)
				(8, 18.21)
				(16, 18.245)
				(32, 17.32)
				(64, 18.697)
			};

		\addplot[fill=green!20, forget plot, opacity=0.8] fill between[of=upper and lower];

		\addplot[thick, green!50!black] coordinates {
				(1, 39.59)
				(2, 33.11)
				(4, 24.08)
				(8, 20.03)
				(16, 18.99)
				(32, 18.36)
				(64, 20.14)
			};
		\addlegendentry{IC3Net}

		\addplot[name path=upper, violet!80!black, forget plot, very thin, draw=none] coordinates {
				(1, 39.98)
				(2, 39.94)
				(4, 38.47)
				(8, 29.19)
				(16, 27.55)
				(32, 26.65)
				(64, 25.27)
			};
		\addplot[name path=lower, violet!80!black, forget plot, very thin, draw=none] coordinates {
				(1, 39.96)
				(2, 39.88)
				(4, 37.01)
				(8, 28.43)
				(16, 26.19)
				(32, 25.03)
				(64, 24.29)
			};

		\addplot[fill=violet!20, forget plot, opacity=0.8] fill between[of=upper and lower];

		\addplot[thick, violet!50!black] coordinates {
				(1, 39.97)
				(2, 39.91)
				(4, 37.74)
				(8, 28.81)
				(16, 26.87)
				(32, 25.85)
				(64, 24.78)
			};
		\addlegendentry{CommNet}

				\addplot[name path=upper, yellow!80!black, forget plot, very thin, draw=none] coordinates {
				(1, 39.70382+0.114/2)
				(2, 31.596+1.44/2)
				(4, 24.644+3.19/2)
				(8, 18.42+1.79/2)
				(16, 17.14+1.95/2)
				(32, 17.350+2.90/2)
				(64, 17.8601+3.46/2)
			};
		\addplot[name path=lower, yellow!80!black, forget plot, very thin, draw=none] coordinates {
				(1, 39.70382-0.114/2)
				(2, 31.596-1.44/2)
				(4, 24.644-3.19/2)
				(8, 18.42-1.79/2)
				(16, 17.14-1.95/2)
				(32, 17.350-2.90/2)
				(64, 17.8601-3.46/2)
			};

		\addplot[fill=yellow!20, forget plot, opacity=0.8] fill between[of=upper and lower];

		\addplot[thick, yellow!75!black] coordinates {
				(1, 39.70382)
				(2, 31.596)
				(4, 24.644)
				(8, 18.42)
				(16, 17.14)
				(32, 17.350)
				(64, 17.8601)
			};
		\addlegendentry{TarMAC}

    		\addplot[name path=upper, red!80!black, forget plot, very thin, draw=none] coordinates {
				(1, 26.86)
				(2, 24.055)
				(4, 19.18)
				(8, 13.20)
				(16, 12.81)
				(32, 12.68)
				(64, 12.65)
			};
		\addplot[name path=lower, red!80!black, forget plot, very thin, draw=none] coordinates {
				(1, 27.47)
				(2, 22.565)
				(4, 16.7)
				(8, 12.975)
				(16, 12.73)
				(32, 12.51)
				(64, 12.48)
			};

		\addplot[fill=red!20, forget plot, opacity=0.8] fill between[of=upper and lower];

		\addplot[ultra thick, red] coordinates {
				(1, 27.17)
				(2, 23.31)
				(4, 17.94)
				(8, 13.09)
				(16, 12.77)
				(32, 12.60)
				(64, 12.57)
			};
		\addlegendentry{\textbf{SLIM}}
	\end{axis}
\end{tikzpicture}}
		\captionsetup{justification=centering}
		\caption{Predator-Prey medium}\label{fig:pp_medium_comm_eff}
	\end{subfigure}
	\begin{subfigure}[b]{0.244\textwidth}
		\centering
		\resizebox{\textwidth}{!}{\begin{tikzpicture}
	\begin{axis}[
			xlabel={Normalised Agent Bandwidth (\(\beta\))},
			ylabel={Success rate (\%)},
			xmode=log,
			xmin=1,
			xmax=64,
			log basis x=2,
			legend pos=north west,
			legend image post style={scale=0.4},
			legend style={
					font=\tiny,
					cells={align=center},
					inner sep=1pt
				},
			width=3.\textwidth,
			height=2.4\textwidth,
			grid=both,
			minor x tick num=1,
			xminorgrids=true,
		]
		\addplot[name path=upper, blue!80!black, forget plot, very thin, draw=none] coordinates {
				(2, 26.35)
				(4, 25.4)
				(8, 34.8)
				(16, 30.5)
				(32, 54.9)
				(64, 91.1)
			};
		\addplot[name path=lower, blue!80!black, forget plot, very thin, draw=none] coordinates {
				(2, 24.65)
				(4, 24.6)
				(8, 17.2)
				(16, 27)
				(32, 34.1)
				(64, 77.9)
			};

		\addplot[fill=blue!20!white, forget plot, opacity=0.8] fill between[of=upper and lower];

		\addplot[thick, blue] coordinates {
				(2, 25.5)
				(4, 25)
				(8, 26)
				(16, 28.75)
				(32, 44.5)
				(64, 84.5)
			};
		\addlegendentry{Commformer (\(1.\))}




		\addplot[name path=upper, green!80!black, forget plot, very thin, draw=none] coordinates {
				(1, 30.0)
				(2, 29.1)
				(4, 37.4)
				(8, 28.9)
				(16, 64.7)
				(32, 80.2)
				(64, 90.1)
			};
		\addplot[name path=lower, green!80!black, forget plot, very thin, draw=none] coordinates {
				(1, 29.1)
				(2, 28.9)
				(4, 29.2)
				(8, 28.5)
				(16, 39.9)
				(32, 55.9)
				(64, 81.3)
			};

		\addplot[fill=green!20!white, forget plot, opacity=0.8] fill between[of=upper and lower];

		\addplot[thick, green!50!black] coordinates {
				(1, 29.5)
				(2, 29.1)
				(4, 33.3)
				(8, 28.7)
				(16, 52.3)
				(32, 68.1)
				(64, 85.7)
			};
		\addlegendentry{IC3Net}

		\addplot[name path=upper, violet!80!black, forget plot, very thin, draw=none] coordinates {
				(1, 22.535)
				(2, 23.095)
				(4, 22.835)
				(8, 22.478)
				(16, 22.725)
				(32, 22.835)
				(64, 41.1)
			};
		\addplot[name path=lower, violet!80!black, forget plot, very thin, draw=none] coordinates {
				(1, 22.115)
				(2, 22.905)
				(4, 22.565)
				(8, 22.322)
				(16, 22.075)
				(32, 22.565)
				(64, 22.4)
			};

		\addplot[fill=violet!20, forget plot, opacity=0.8] fill between[of=upper and lower];

		\addplot[thick, violet!50!black] coordinates {
				(1, 22.3)
				(2, 23.0)
				(4, 22.6)
				(8, 22.5)
				(16, 22.4)
				(32, 22.7)
				(64, 31.9)
			};
		\addlegendentry{CommNet}

		\addplot[name path=upper, yellow!80!black, forget plot, very thin, draw=none] coordinates {
				(1, 30.163 + .88/2)
				(2, 29.469 + .80/2)
				(4, 28.894 + .57/2)
				(8, 28.919 + .31/2)
				(16, 37.481 + 16.0/2)
				(32, 40.925 + 23.23/2)
				(64, 65.488 + 26.19/2)
			};
		\addplot[name path=lower, yellow!80!black, forget plot, very thin, draw=none] coordinates {
				(1, 30.163 - .88/2)
				(2, 29.469 - .80/2)
				(4, 28.894 - .57/2)
				(8, 28.919 - .31/2)
				(16, 37.481 - 16.0/2)
				(32, 40.925 - 23.23/2)
				(64, 65.488 - 26.19/2)
			};

		\addplot[fill=yellow!20, forget plot, opacity=0.8] fill between[of=upper and lower];

		\addplot[thick, yellow!75!black] coordinates {
				(1, 30.163)
				(2, 29.469)
				(4, 28.894)
				(8, 28.919)
				(16, 37.481)
				(32, 40.925)
				(64, 65.488)
			};
		\addlegendentry{TarMAC}

		\addplot[name path=upper, red!80!black, forget plot, very thin, draw=none] coordinates {
				(1, 24.2)
				(2, 24.3)
				(4, 74.2)
				(8, 99.02)
				(16, 99.31)
				(32, 99.51)
				(64, 99.4)
			};
		\addplot[name path=lower, red!80!black, forget plot, very thin, draw=none] coordinates {
				(1, 24.2)
				(2, 24.3)
				(4, 56)
				(8, 98.18)
				(16, 98.09)
				(32, 98.99)
				(64, 99.1)
			};

		\addplot[fill=red!20, forget plot, opacity=0.8] fill between[of=upper and lower];

		\addplot[ultra thick, red] coordinates {
				(1, 24.2)
				(2, 24.3)
				(4, 65.1)
				(8, 98.6)
				(16, 98.7)
				(32, 99.3)
				(64, 99.3)
			};
		\addlegendentry{\textbf{SLIM}}

	\end{axis}
\end{tikzpicture}}
		\captionsetup{justification=centering}
		\caption{Traffic Junction easy}\label{fig:tj_easy_comm_eff}
	\end{subfigure}
	\begin{subfigure}[b]{0.25\textwidth}
		\centering
		\resizebox{\textwidth}{!}{\begin{tikzpicture}
	\begin{axis}[
			xlabel={Normalised Agent Bandwidth (\(\beta\))},
			ylabel={Success rate (\%)},
			xmode=log,
			xmin=1,
			xmax=64,
			log basis x=2,
			legend pos=north west,
			legend image post style={scale=0.4},
			legend style={
					font=\tiny,
					cells={align=center},
					inner sep=1pt
				},
			width=3.\textwidth,
			height=2.4\textwidth,
			grid=both,
			minor x tick num=1,
			xminorgrids=true,
		]
		\addplot[name path=upper, blue!80!black, forget plot, very thin, draw=none] coordinates {
				(2, 80.7)
				(4, 81.5)
				(8, 78.35)
				(16, 82.3)
				(32, 99)
				(64, 98.37)
			};
		\addplot[name path=lower, blue!80!black, forget plot, very thin, draw=none] coordinates {
				(2, 79.3)
				(4, 79.5)
				(8, 76.65)
				(16, 73.2)
				(32, 91.5)
				(64, 93.7)
			};

		\addplot[fill=blue!20!white, forget plot, opacity=0.8] fill between[of=upper and lower];

		\addplot[thick, blue] coordinates {
				(2, 80)
				(4, 80.5)
				(8, 77.5)
				(16, 77.75)
				(32, 95.25)
				(64, 96.0)
			};
		\addlegendentry{Commformer (\(1.\))}




		\addplot[name path=upper, green!80!black, forget plot, very thin, draw=none] coordinates {
				(1, 73.05)
				(2, 72.5)
				(4, 73.9)
				(8, 73.4)
				(16, 73.8)
				(32, 77.6)
				(64, 84.3)
			};
		\addplot[name path=lower, green!80!black, forget plot, very thin, draw=none] coordinates {
				(1, 72.95)
				(2, 69.1)
				(4, 73.5)
				(8, 72.6)
				(16, 72.6)
				(32, 73.8)
				(64, 76.9)
			};

		\addplot[fill=green!20!white, forget plot, opacity=0.8] fill between[of=upper and lower];

		\addplot[thick, green!50!black] coordinates {
				(1, 73.0)
				(2, 70.8)
				(4, 73.7)
				(8, 73.0)
				(16, 73.2)
				(32, 75.6)
				(64, 80.6)
			};
		\addlegendentry{IC3Net}

		\addplot[name path=upper, violet!80!black, forget plot, very thin, draw=none] coordinates {
				(1, 72.53)
				(2, 72.80)
				(4, 74.21)
				(8, 73.24)
				(16, 73.7)
				(32, 72.6)
				(64, 79.6)
			};
		\addplot[name path=lower, violet!80!black, forget plot, very thin, draw=none] coordinates {
				(1, 61.23)
				(2, 72.34)
				(4, 73.41)
				(8, 72.36)
				(16, 72.7)
				(32, 71.8)
				(64, 74.4)
			};

		\addplot[fill=violet!20, forget plot, opacity=0.8] fill between[of=upper and lower];

		\addplot[thick, violet!50!black] coordinates {
				(1, 66.89)
				(2, 72.57)
				(4, 73.81)
				(8, 72.82)
				(16, 73.20)
				(32, 72.26)
				(64, 77.0)
			};
		\addlegendentry{CommNet}

		\addplot[name path=upper, yellow!80!black, forget plot, very thin, draw=none] coordinates {
				(1, 71.00 + 1.55/2)
				(2, 72.96 + 1.31/2)
				(4, 73.27 + .88/2)
				(8, 72.96 + .50/2)
				(16, 73.27 + .49/2)
				(32, 72.76 + 1.48/2)
				(64, 70.95 + 2.8/2)
			};
		\addplot[name path=lower, yellow!80!black, forget plot, very thin, draw=none] coordinates {
				(1, 71.00 - 1.55/2)
				(2, 72.96 - 1.31/2)
				(4, 73.27 - .88/2)
				(8, 72.96 - .50/2)
				(16, 73.27 - .49/2)
				(32, 72.76 - 1.48/2)
				(64, 70.95 - 2.8/2)
			};

		\addplot[fill=yellow!20, forget plot, opacity=0.8] fill between[of=upper and lower];

		\addplot[thick, yellow!75!black] coordinates {
				(1, 71.00)
				(2, 72.96)
				(4, 73.27)
				(8, 72.96)
				(16, 73.27)
				(32, 72.76)
				(64, 70.95)
			};
		\addlegendentry{TarMAC}

		\addplot[name path=upper, red!80!black, forget plot, very thin, draw=none] coordinates {
				(1, 74.1)
				(2, 68.4)
				(4, 96.15)
				(8, 99.65)
				(16, 99.88)
				(32, 99.36)
				(64, 97.35)
			};
		\addplot[name path=lower, red!80!black, forget plot, very thin, draw=none] coordinates {
				(1, 70.3)
				(2, 66.7)
				(4, 90.05)
				(8, 98.15)
				(16, 99.72)
				(32, 98.63)
				(64, 96.94)
			};

		\addplot[fill=red!20, forget plot, opacity=0.8] fill between[of=upper and lower];

		\addplot[ultra thick, red] coordinates {
				(1, 72.2)
				(2, 67.6)
				(4, 92.5)
				(8, 99.3)
				(16, 99.8)
				(32, 99.0)
				(64, 97.2)
			};
		\addlegendentry{\textbf{SLIM}}

	\end{axis}
\end{tikzpicture}}
		\captionsetup{justification=centering}
		\caption{Traffic Junction medium}\label{fig:tj_medium_comm_eff}
	\end{subfigure}
	\caption{\textbf{Performance of SLIM and baselines across a logarithmic range of normalised agent bandwidth values $\beta$ (from $2^0$ to $2^6$).} Top: mean episode length for the Predator-Prey environment (lower values indicate better performance). Bottom: success rate for the Traffic Junction environment (higher is better). Shaded regions denote the standard error of the mean across 4 seeds. One can note that data points for the dense variant of CommFormer are absent in the lowest bandwidth regimes ($\beta = 2^0$) because the architectural constraints of the model prevent it from satisfying such strict bandwidth limits. See details in~\Cref{section:bandwidth_hyp}.} 
	\label{fig:comm_eff_results}
\end{figure*}

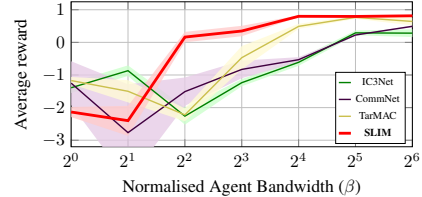
\begin{wrapfigure}{r}{0.4\textwidth}
		\centering
		\resizebox{\linewidth}{!}{\begin{tikzpicture}
	\begin{axis}[
			xlabel={Normalised Agent Bandwidth (\(\beta\))},
			ylabel={Average reward},
			xmode=log,
			xmin=1,
			xmax=64,
			ymin=-3.2,
			log basis x=2,
			legend image post style={scale=0.4},
			legend style={
					at={(.98, .02)},
					anchor=south east,
					font=\tiny,
					cells={align=center},
					inner sep=1pt
				},
			width=1.5\linewidth,
			height=.8\linewidth,
			grid=both,
			minor x tick num=1,
			xminorgrids=true,
		]






		\addplot[name path=upper, green!80!black, forget plot, very thin, draw=none] coordinates {
				(1, -1.3916 + .32207/2)
				(2, -.86812 + .32458/2)
				(4, -2.26121 + .522/2)
				(8, -1.23597 + .21771/2)
				(16, -.61101 + .21339/2)
				(32, .29082 + .14536/2)
				(64, .28252 + .24078/2)
			};
		\addplot[name path=lower, green!80!black, forget plot, very thin, draw=none] coordinates {
				(1, -1.3916 - .32207/2)
				(2, -.86812 - .32458/2)
				(4, -2.26121 - .522/2)
				(8, -1.23597 - .21771/2)
				(16, -.61101 - .21339/2)
				(32, .29082 - .14536/2)
				(64, .28252 - .24078/2)
			};

		\addplot[fill=green!20, forget plot, opacity=0.8] fill between[of=upper and lower];

		\addplot[thick, green!50!black] coordinates {
				(1, -1.3916)
				(2, -.86812)
				(4, -2.26121)
				(8, -1.23597)
				(16, -.61101)
				(32, .29082)
				(64, .28252)
			};
		\addlegendentry{IC3Net}

		\addplot[name path=upper, violet!80!black, forget plot, very thin, draw=none] coordinates {
				(1, -1.24897 + 1.33951/2)
				(2, -2.76744 + 2.96218/2)
				(4, -1.51017 + .86706/2)
				(8, -.82918 + .53939/2)
				(16, -.53104 + .15562/2)
				(32, .2254 + .045/2)
				(64, .4918 + .057/2)
			};
		\addplot[name path=lower, violet!80!black, forget plot, very thin, draw=none] coordinates {
				(1, -1.24897 - 1.33951/2)
				(2, -2.76744 - 2.96218/2)
				(4, -1.51017 - .86706/2)
				(8, -.82918 - .53939/2)
				(16, -.53104 - .15562/2)
				(32, .2254 - .045/2)
				(64, .4918 - .057/2)
			};

		\addplot[fill=violet!20, forget plot, opacity=0.8] fill between[of=upper and lower];

		\addplot[thick, violet!50!black] coordinates {
				(1, -1.24897)
				(2, -2.76744)
				(4, -1.51017)
				(8, -.82918)
				(16, -.53104)
				(32, .2254)
				(64, .4918)
			};
		\addlegendentry{CommNet}

		\addplot[name path=upper, yellow!80!black, forget plot, very thin, draw=none] coordinates {
				(1, -1.1708 + .2211/2)
				(2, -1.49953 + .68065/2)
				(4, -2.22172 + .39406/2)
				(8, -.46964 + .71314/2)
				(16, .49058 + .15521/2)
				(32, .76666 + .085612/2)
				(64, .64117 + .24307/2)
			};
		\addplot[name path=lower, yellow!80!black, forget plot, very thin, draw=none] coordinates {
				(1, -1.1708 - .2211/2)
				(2, -1.49953 - .68065/2)
				(4, -2.22172 - .39406/2)
				(8, -.46964 - .71314/2)
				(16, .49058 - .15521/2)
				(32, .76666 - .085612/2)
				(64, .64117 - .24307/2)
			};

		\addplot[fill=yellow!20, forget plot, opacity=0.8] fill between[of=upper and lower];

		\addplot[thick, yellow!75!black] coordinates {
				(1, -1.1708)
				(2, -1.49953)
				(4, -2.22172)
				(8, -.46964)
				(16, .49058)
				(32, .76666)
				(64, .64117)
			};
		\addlegendentry{TarMAC}

		\addplot[name path=upper, red!80!black, forget plot, very thin, draw=none] coordinates {
				(1, -1.9799)
				(2, -1.9466)
				(4, .3213)
				(8, .5074)
				(16, .82612)
				(32, .8393)
				(64, .8401)
			};
		\addplot[name path=lower, red!80!black, forget plot, very thin, draw=none] coordinates {
				(1, -2.2905)
				(2, -2.865)
				(4, .0033)
				(8, .188)
				(16, .77152)
				(32, .75793)
				(64, .78881)
			};

		\addplot[fill=red!20, forget plot, opacity=0.8] fill between[of=upper and lower];

		\addplot[ultra thick, red] coordinates {
				(1, -2.135)
				(2, -2.40)
				(4, .1623)
				(8, .348)
				(16, .79882)
				(32, .79443)
				(64, .81441)
			};
		\addlegendentry{\textbf{SLIM}}

	\end{axis}
\end{tikzpicture}}
		\caption{\textbf{Performance on a logarithmic range of normalised agent bandwidth values $\beta$ in the Navigation environment.} Shaded regions denote the standard error. SLIM achieves the best performance in high bandwidth settings and is more resilient to the reduction of the bandwidth.}\label{fig:nav_comm_eff}
	\end{wrapfigure}

To evaluate the robustness of each architecture, we swept the normalised agent bandwidth $\beta$ across a logarithmic scale ranging from $2^0$ to $2^6$. For each value, we trained every model on 4 distinct random seeds, resulting in a total of 28 training runs per model. \Cref{fig:comm_eff_results} and \Cref{fig:nav_comm_eff} illustrate the evolution of performance as a function of $\beta$, with shaded regions indicating the standard error.
CommFormer achieves competitive performance in high-bandwidth regimes; however, this performance degrades significantly as the bandwidth constraint tightens. The sparse variant ($\sigma = 0.5$) exhibits higher variability and lower results. We excluded this variant from Traffic Junction because a fixed sparse communication graph is unsuitable for tasks where specific and dynamic interactions are required (e.g., cars at the same intersection). Similarly, IC3Net and CommNet are less robust, showing either consistently lower returns or quick collapse under limited bandwidth. We observe similar trends for TarMAC, except in Navigation where it maintains a more stable performance.
In contrast, SLIM demonstrates robustness under bandwidth constraints. It matches or surpasses the peak performance of the strongest baseline at high bandwidth while maintaining high rewards even under strict constraints where others fail. While a minor performance drop occurs in Traffic Junction under severe bandwidth limitations, SLIM consistently offers robust performances. It validates its ability to learn robust policies without relying on excessive transmission volume.

\subsection{Ablation Study: Impact of Message History Cache}
\label{sec:ablation_cache}

\begin{figure*}
	\centering
	\begin{subfigure}[t]{.245\linewidth}
		\begingroup
		\centering
		\begin{tikzpicture}[baseline]
			\tiny
			\definecolor{darkgray176}{RGB}{176,176,176}
			\definecolor{darkorange25512714}{RGB}{255,0,0}
			\definecolor{forestgreen4416044}{RGB}{44,160,44}
			\definecolor{lightgray204}{RGB}{204,204,204}
			\definecolor{steelblue31119180}{RGB}{100,0,255}

			\begin{axis}[
                    tick style={line width=0.3pt, major tick length=2pt, minor tick length=1pt},
					xticklabel style={font=\fontsize{4}{5}\selectfont},
					yticklabel style={font=\fontsize{4}{5}\selectfont},
					legend cell align={left},
					legend style={fill opacity=0.8, draw opacity=1, text opacity=1, draw=lightgray204},
                    legend image post style={xscale=.25},
					tick align=outside,
					tick pos=left,
					width=1.15\linewidth,
					height=1.\linewidth,
					xlabel={Epoch},
					xmajorgrids,
					ylabel={Steps},
					ymajorgrids,
					legend columns=1,
					legend pos=north east,
					xticklabels={,0,,1000,, 2000},
					yticklabels={,,5,5.5,6},
					ymax=6,
					xmin=-0.1,
					xmax=2000,
				]
				\addplot [thick, darkorange25512714]
				table[y index=1,  x expr=\coordindex*10, col sep=comma] {figures/data/easy_8_no_cache.csv};

				\addplot [name path=upper, draw=none, forget plot]
				table [y index=2, x expr=\coordindex*10, col sep=comma] {figures/data/easy_8_no_cache.csv};

				\addplot [name path=lower, draw=none, forget plot]
				table [y index=3, x expr=\coordindex*10, col sep=comma] {figures/data/easy_8_no_cache.csv};

				\addplot [fill=darkorange25512714!20, forget plot] fill between[of=upper and lower];
				\addlegendentry{w/o Cache};

				\addplot [thick, steelblue31119180]
				table[y index=1,  x expr=\coordindex*10, col sep=comma] {figures/data/easy_8_cache.csv};
				\addlegendentry{w/ Cache};

				\addplot [name path=upper, draw=none]
				table [y index=2, x expr=\coordindex*10, col sep=comma] {figures/data/easy_8_cache.csv};

				\addplot [name path=lower, draw=none]
				table [y index=3, x expr=\coordindex*10, col sep=comma] {figures/data/easy_8_cache.csv};

				\addplot [fill=steelblue31119180!20] fill between[of=upper and lower];


			\end{axis}
		\end{tikzpicture}
		\caption{Low bandwidth}\label{pp-low}
		\endgroup
	\end{subfigure}
	\begin{subfigure}[t]{.245\linewidth}
		\begingroup
		\centering
		\begin{tikzpicture}[baseline]
			\tiny
			\definecolor{darkgray176}{RGB}{176,176,176}
			\definecolor{darkorange25512714}{RGB}{255,0,0}
			\definecolor{forestgreen4416044}{RGB}{44,160,44}
			\definecolor{lightgray204}{RGB}{204,204,204}
			\definecolor{steelblue31119180}{RGB}{100,0,255}

			\begin{axis}[
                    tick style={line width=0.3pt, major tick length=2pt, minor tick length=1pt},
					xticklabel style={font=\fontsize{4}{5}\selectfont},
					yticklabel style={font=\fontsize{4}{5}\selectfont},
					legend cell align={left},
					legend style={fill opacity=0.8, draw opacity=1, text opacity=1, draw=lightgray204},
                    legend image post style={xscale=.25},
					tick align=outside,
					tick pos=left,
					width=1.15\linewidth,
					height=1.\linewidth,
					xlabel={Epoch},
					xmajorgrids,
					ylabel={Steps},
					ymajorgrids,
					legend columns=1,
					legend pos=north east,
					xticklabels={,0,,1000,, 2000},
					yticklabels={,,5,5.5,6},
					ymax=6,
					xmin=-0.1,
					xmax=2000,
				]
				\addplot [thick, darkorange25512714]
				table[y index=1,  x expr=\coordindex*10, col sep=comma] {figures/data/easy_64_no_cache.csv};

				\addplot [name path=upper, draw=none, forget plot]
				table [y index=2, x expr=\coordindex*10, col sep=comma] {figures/data/easy_64_no_cache.csv};

				\addplot [name path=lower, draw=none, forget plot]
				table [y index=3, x expr=\coordindex*10, col sep=comma] {figures/data/easy_64_no_cache.csv};

				\addplot [fill=darkorange25512714!20, forget plot] fill between[of=upper and lower];
				\addlegendentry{w/o Cache};

				\addplot [thick, steelblue31119180]
				table[y index=1,  x expr=\coordindex*10, col sep=comma] {figures/data/easy_64_cache.csv};

				\addplot [name path=upper, draw=none]
				table [y index=2, x expr=\coordindex*10, col sep=comma] {figures/data/easy_64_cache.csv};

				\addplot [name path=lower, draw=none]
				table [y index=3, x expr=\coordindex*10, col sep=comma] {figures/data/easy_64_cache.csv};

				\addplot [fill=steelblue31119180!20] fill between[of=upper and lower];
				\addlegendentry{w/ Cache};


			\end{axis}
		\end{tikzpicture}
		\caption{High bandwidth}\label{pp-high}
		\endgroup
	\end{subfigure}
	\begin{subfigure}[t]{.245\linewidth}
		\begingroup
		\centering
		\begin{tikzpicture}[baseline]
			\tiny
			\definecolor{darkgray176}{RGB}{176,176,176}
			\definecolor{darkorange25512714}{RGB}{255,0,0}
			\definecolor{forestgreen4416044}{RGB}{44,160,44}
			\definecolor{lightgray204}{RGB}{204,204,204}
			\definecolor{steelblue31119180}{RGB}{100,0,255}

			\begin{axis}[
                    tick style={line width=0.3pt, major tick length=2pt, minor tick length=1pt},
					xticklabel style={font=\fontsize{4}{5}\selectfont},
					yticklabel style={font=\fontsize{4}{5}\selectfont},
					legend cell align={left},
					legend style={fill opacity=0.8, draw opacity=1, text opacity=1, draw=lightgray204},
                    legend image post style={xscale=.25},
					tick align=outside,
					tick pos=left,
					width=1.15\linewidth,
					height=1.\linewidth,
					xlabel={Epoch},
					xmajorgrids,
					ylabel={Reward},
					ymajorgrids,
					legend columns=1,
					legend pos=south east,
					xticklabels={,0,,1000,, 2000},
					yticklabels={,-0.2, -0.1,0},
					ymax=0,
					xmin=-0.1,
					xmax=2000,
				]
				\addplot [thick, darkorange25512714]
				table[y index=1,  x expr=\coordindex*10, col sep=comma] {figures/data/shapes_8_no_cache.csv};

				\addplot [name path=upper, draw=none, forget plot]
				table [y index=2, x expr=\coordindex*10, col sep=comma] {figures/data/shapes_8_no_cache.csv};

				\addplot [name path=lower, draw=none, forget plot]
				table [y index=3, x expr=\coordindex*10, col sep=comma] {figures/data/shapes_8_no_cache.csv};

				\addplot [fill=darkorange25512714!20, forget plot] fill between[of=upper and lower];
				\addlegendentry{w/o Cache};

				\addplot [thick, steelblue31119180]
				table[y index=1,  x expr=\coordindex*10, col sep=comma] {figures/data/shapes_8_cache.csv};

				\addplot [name path=upper, draw=none]
				table [y index=2, x expr=\coordindex*10, col sep=comma] {figures/data/shapes_8_cache.csv};

				\addplot [name path=lower, draw=none]
				table [y index=3, x expr=\coordindex*10, col sep=comma] {figures/data/shapes_8_cache.csv};

				\addplot [fill=steelblue31119180!20] fill between[of=upper and lower];
				\addlegendentry{w/ Cache};


			\end{axis}
		\end{tikzpicture}
		\caption{Low bandwidth}\label{shapes-low}
		\endgroup
	\end{subfigure}
	\begin{subfigure}[t]{.245\linewidth}
		\begingroup
		\centering
		\begin{tikzpicture}[baseline]
			\tiny
			\definecolor{darkgray176}{RGB}{176,176,176}
			\definecolor{darkorange25512714}{RGB}{255,0,0}
			\definecolor{forestgreen4416044}{RGB}{44,160,44}
			\definecolor{lightgray204}{RGB}{204,204,204}
			\definecolor{steelblue31119180}{RGB}{100,0,255}

			\begin{axis}[
                    tick style={line width=0.3pt, major tick length=2pt, minor tick length=1pt},
					xticklabel style={font=\fontsize{4}{5}\selectfont},
					yticklabel style={font=\fontsize{4}{5}\selectfont},
					legend cell align={left},
					legend style={fill opacity=0.8, draw opacity=1, text opacity=1, draw=lightgray204},
                    legend image post style={xscale=.25},
					tick align=outside,
					tick pos=left,
					width=1.15\linewidth,
					height=1.\linewidth,
					xlabel={Epoch},
					xmajorgrids,
					ylabel={Reward},
					ymajorgrids,
					legend columns=1,
					legend pos=south east,
					xticklabels={,0,,1000,, 2000},
					yticklabels={,-0.2, -0.1,0},
					ymax=0,
					xmin=-0.1,
					xmax=2000,
				]
				\addplot [thick, darkorange25512714]
				table[y index=1,  x expr=\coordindex*10, col sep=comma] {figures/data/shapes_64_no_cache.csv};

				\addplot [name path=upper, draw=none, forget plot]
				table [y index=2, x expr=\coordindex*10, col sep=comma] {figures/data/shapes_64_no_cache.csv};

				\addplot [name path=lower, draw=none, forget plot]
				table [y index=3, x expr=\coordindex*10, col sep=comma] {figures/data/shapes_64_no_cache.csv};

				\addplot [fill=darkorange25512714!20, forget plot] fill between[of=upper and lower];
				\addlegendentry{w/o Cache};

				\addplot [thick, steelblue31119180]
				table[y index=1,  x expr=\coordindex*10, col sep=comma] {figures/data/shapes_64_cache.csv};

				\addplot [name path=upper, draw=none]
				table [y index=2, x expr=\coordindex*10, col sep=comma] {figures/data/shapes_64_cache.csv};

				\addplot [name path=lower, draw=none]
				table [y index=3, x expr=\coordindex*10, col sep=comma] {figures/data/shapes_64_cache.csv};

				\addplot [fill=steelblue31119180!20] fill between[of=upper and lower];
				\addlegendentry{w/ Cache};


			\end{axis}
		\end{tikzpicture}
		\caption{High bandwidth}\label{shapes-high}
		\endgroup
	\end{subfigure}
	\caption{\textbf{Ablation study on the effect of the cache on an non jointly observable environment.} We compare the performance of SLIM with and without the temporal cache mechanism in the Predator-Prey easy environment {((\ref{pp-low}) and (\ref{pp-high})) and the SHAPES environment ((\ref{shapes-low}) and (\ref{shapes-high}))} under two different communication bandwidths: \(2^3\) ({ \ref{pp-low}) and (\ref{shapes-low})}) and \(2^6\) ((\ref{pp-high}) {and (\ref{shapes-high})}). The results demonstrate that the cache significantly improves results along the training process while increasing stability{; for non-jointly fully observable environments}. The line reported is the mean over 4 seeds, with the shaded area representing the standard error.}
	\label{fig:abla_graphs}
\end{figure*}
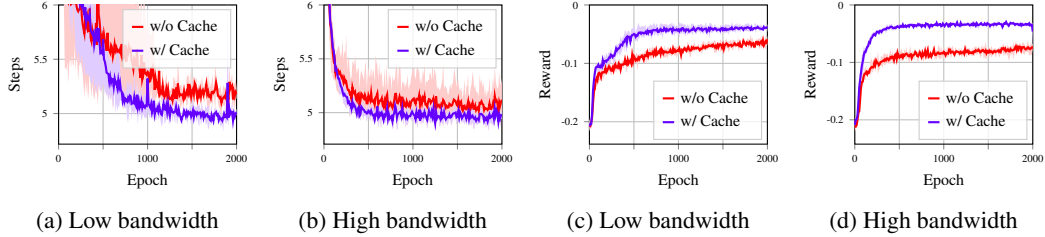

To validate the hypothesis that the historical context is beneficial for environments with partial observability, we conducted an ablation study comparing the full SLIM architecture against the variant with the cache deactivated. The results, presented in~\Cref{fig:abla_graphs}, demonstrate the clear advantage of the memory mechanism. Additional detailed results are available in~\Cref{tab:ablation_cache}. In the Predator-Prey and the SHAPES environments, where tracking agent trajectories over time is advantageous, the cache-enabled model consistently outperforms the cache-disabled baseline, achieving the objective in fewer steps in Predator-Prey and higher rewards in SHAPES. The only exception occurs at very low bandwidths in Predator-Prey, where we hypothesise that the limited representational capacity of the latent space prevents the effective disentanglement of spatial features, temporal dependencies, and agent identities.

\setlength{\mycolspace}{2.5pt}

\section{Limitations}

Our study targets algorithmic communication efficiency and does not constitute a complete model of real wireless deployment. The normalised bandwidth budget \(\beta\) provides a controlled proxy unifying message dimension, communication rounds, and graph sparsity, but it abstracts away packet headers, quantisation, latency, routing overhead, packet loss, and medium contention. Extending \(\beta\) to capture some of these factors, and evaluating SLIM under more realistic network conditions, is left for future work. We also do not empirically combine SLIM with information-theoretic message compression, which we hypothesise to be complementary but leave to future work. The message history cache had negligible overhead in our experiments, but its memory cost grows linearly with episode length; windowing strategies or compressed memory representations could be considered for \textit{very} long-horizon tasks. 

\section{Conclusion}

We proposed SLIM, a simple and modular approach to communication in multi-agent reinforcement learning that explicitly decouples the architecture responsible for inter-agent messaging encoding, from that dedicated to policy execution (responsible for actions). We introduced an evaluation protocol based on a normalised agent bandwidth metric (\(\beta\)). This unified measure integrates message size, transmission frequency, and graph sparsity into a single constraint, enabling a systematic benchmarking of diverse communication strategies under identical physical limitations. Our empirical results across multiple partially observable benchmarks demonstrate that SLIM not only matches state-of-the-art methods in high-bandwidth regimes but exhibits clear robustness as constraints tighten. While baseline performance degrade rapidly under severe constraints, SLIM maintains effective coordination even with lower bandwidth limits. Finally, ablation studies confirmed the benefit of our proposed message history cache for improving learning in non-jointly observable environments.
Future work will further explore connections between modern communication-efficient MARL settings and complementary developments in network science, where transmission rate and information-theoretical views have been widely considered; perhaps allowing further calibration of communication efficiency in a more general sense.

\section*{Acknowledgements}

This work received financial support from Crédit Agricole SA through the research chair  \textit{Trustworthy and Responsible AI} at École Polytechnique.

This work was granted access to the HPC resources of IDRIS under the allocation 2025-AD011017102 made by GENCI.

Finally, we would like to thank Mathis Le Bail, Clément Elliker and Mahammed Elsharkawy for their help and insights throughout the project.

\bibliographystyle{plain}
\bibliography{biblio}

\newpage
\appendix
\section{Experiment Details}

\newcommand{\lrvalue}{\mbox{\(5{\times}10^{-4}\)}}

\begin{table}[h]
	\caption{\textbf{Hyperparameter configuration for the SLIM architecture across all benchmarks.}
		These values were optimised using a fixed message dimensionality of $d=64$ ($2^6$) to ensure that the architecture remains robust across varying bandwidths without overfitting to specific constraints.}
	\label{tab:slim_hyperparams}
	\centering
	\setlength{\tabcolsep}{3.5pt}
	\begin{tabularx}{\textwidth}{c|CC|CC|C|C}
		\toprule
		                                 & \multicolumn{2}{c}{Predator-Prey} & \multicolumn{2}{|c|}{Traffic Junction} & Navigation          & SHAPES                                                          \\
		\midrule
		                                 & Easy                              & Medium                                 & Easy                & Medium              & /                   & /                   \\
		\midrule
		Clip param. \(\varepsilon\)      & 0.2                               & 0.2                                    & 0.2                 & 0.2                 & 0.2                 & 0.2                 \\
		Entropy coeff. \(\sigma\)        & 0.02                              & 0.02                                   & 0.02                & 0.02                & 0.02                & 0.02                \\
		Episodes/epoch                   & 500                               & 500                                    & 500                 & 500                 & 100                 & 100                 \\
		Number of epochs                 & 2000                              & 2000                                   & 600                 & 4000                & 3200                & 2000                \\
		PPO Epochs                       & 5                                 & 5                                      & 1                   & 1                   & 1                   & 1                   \\
		Hidden size                      & 128                               & 128                                    & 128                 & 128                 & 128                 & 128                 \\
		Message dim.                     & \(2^6\)                           & \(2^6\)                                & \(2^6\)             & \(2^6\)             & \(2^6\)             & \(2^6\)             \\
		Using cache                      & \texttt{True}                     & \texttt{True}                          & \texttt{False}      & \texttt{False}      & \texttt{False}      & \texttt{True}       \\
		Discount \(\gamma\)              & 0.99                              & 0.99                                   & 0.99                & 0.99                & 0.99                & 0.99                \\
		Learning rate                    & \lrvalue                          & \lrvalue                               & \lrvalue            & \lrvalue            & \lrvalue            & \lrvalue            \\
		Replay buffer (\# ep.)           & 1000                              & 1000                                   & 1000                & 1000                & 250                 & 5000                \\
		GAE \(\lambda\)                  & 0.95                              & 0.95                                   & 0.95                & 0.95                & 0.95                & 0.95                \\
		\bottomrule
	\end{tabularx}
\end{table}

\subsection{Detailed Communication Parameters}
\label{section:bandwidth_hyp}

\begin{table}[h]
	\caption{\textbf{Communication parameters for the normalized agent bandwidth configurations.}
	For each bandwidth budget \(\beta\), we report the graph density \(\sigma\), communication rounds \(k\), and largest feasible message dimension \(d\) satisfying \(\sigma \times k \times d \leq \beta\). TarMAC is evaluated in the dense one-pass setting used in our sweeps. Dense CommFormer cannot satisfy \(\beta=2^0\) with integer \(d\geq1\). En-dashes (--) denote inaccessible configurations.}
	\label{tab:communication_parameters}
	\centering
	\setlength{\tabcolsep}{1.0pt}
	\renewcommand{\arraystretch}{1.08}
	\begin{tabular*}{\textwidth}{@{\extracolsep{\fill}}l*{7}{ccc}@{}}
		\toprule
		& \multicolumn{3}{c}{\(\beta=2^0\)}
		& \multicolumn{3}{c}{\(\beta=2^1\)}
		& \multicolumn{3}{c}{\(\beta=2^2\)}
		& \multicolumn{3}{c}{\(\beta=2^3\)}
		& \multicolumn{3}{c}{\(\beta=2^4\)}
		& \multicolumn{3}{c}{\(\beta=2^5\)}
		& \multicolumn{3}{c}{\(\beta=2^6\)} \\
		\cmidrule(lr){2-4}
		\cmidrule(lr){5-7}
		\cmidrule(lr){8-10}
		\cmidrule(lr){11-13}
		\cmidrule(lr){14-16}
		\cmidrule(lr){17-19}
		\cmidrule(l){20-22}
		Model
		& \(\sigma\) & \(k\) & \(d\)
		& \(\sigma\) & \(k\) & \(d\)
		& \(\sigma\) & \(k\) & \(d\)
		& \(\sigma\) & \(k\) & \(d\)
		& \(\sigma\) & \(k\) & \(d\)
		& \(\sigma\) & \(k\) & \(d\)
		& \(\sigma\) & \(k\) & \(d\) \\
		\midrule
		CommFormer
		& -- & -- & --
		& 1.0 & 2 & \(2^0\)
		& 1.0 & 2 & \(2^1\)
		& 1.0 & 2 & \(2^2\)
		& 1.0 & 2 & \(2^3\)
		& 1.0 & 2 & \(2^4\)
		& 1.0 & 2 & \(2^5\) \\
		CommFormer (sparse)
		& 0.5 & 2 & \(2^0\)
		& 0.5 & 2 & \(2^1\)
		& 0.5 & 2 & \(2^2\)
		& 0.5 & 2 & \(2^3\)
		& 0.5 & 2 & \(2^4\)
		& 0.5 & 2 & \(2^5\)
		& 0.5 & 2 & \(2^6\) \\
		CommNet
		& 1.0 & 1 & \(2^0\)
		& 1.0 & 1 & \(2^1\)
		& 1.0 & 1 & \(2^2\)
		& 1.0 & 1 & \(2^3\)
		& 1.0 & 1 & \(2^4\)
		& 1.0 & 1 & \(2^5\)
		& 1.0 & 1 & \(2^6\) \\
		IC3Net
		& 1.0 & 1 & \(2^0\)
		& 1.0 & 1 & \(2^1\)
		& 1.0 & 1 & \(2^2\)
		& 1.0 & 1 & \(2^3\)
		& 1.0 & 1 & \(2^4\)
		& 1.0 & 1 & \(2^5\)
		& 1.0 & 1 & \(2^6\) \\
		TarMAC
		& 1.0 & 1 & \(2^0\)
		& 1.0 & 1 & \(2^1\)
		& 1.0 & 1 & \(2^2\)
		& 1.0 & 1 & \(2^3\)
		& 1.0 & 1 & \(2^4\)
		& 1.0 & 1 & \(2^5\)
		& 1.0 & 1 & \(2^6\) \\
		SLIM
		& 1.0 & 1 & \(2^0\)
		& 1.0 & 1 & \(2^1\)
		& 1.0 & 1 & \(2^2\)
		& 1.0 & 1 & \(2^3\)
		& 1.0 & 1 & \(2^4\)
		& 1.0 & 1 & \(2^5\)
		& 1.0 & 1 & \(2^6\) \\
		\bottomrule
	\end{tabular*}
\end{table}

\section{Additional Results}

\newcommand{\ablval}[2]{%
	#1{\scriptsize\(\pm #2\)}%
}
\newcommand{\ablbest}[2]{%
	\textbf{#1{\scriptsize\(\pm #2\)}}%
}

\begin{table}[t]
	\centering
	\caption{\textbf{Results of the ablation study on the impact of the cache in non-jointly observable environments.} \label{tab:ablation_cache}
		We report mean episode length across varying communication dimensions for Predator-Prey and rewards for SHAPES.
		Lower values are better for episode length; higher values are better for rewards.}
	\setlength{\tabcolsep}{2pt}
	\begin{tabular*}{\textwidth}{@{\extracolsep{\fill}}llccccccc@{}}
		\toprule
		Setting & Cache & \multicolumn{7}{c}{\(\beta\)} \\
		\midrule
		        &       & 1 & 2 & 4 & 8 & 16 & 32 & 64 \\
		\midrule
		\multirow{2}{*}{Pred.-Prey Easy}   & w/  & \ablbest{6.61}{.29}  & \ablval{7.10}{1.2}   & \ablbest{5.19}{.18}  & \ablbest{5.01}{.03}  & \ablbest{4.99}{.03}  & \ablbest{4.99}{.04}  & \ablbest{4.97}{.04} \\
		                                   & w/o & \ablbest{6.65}{.34}  & \ablbest{6.49}{.1}   & \ablbest{5.16}{.06}  & \ablval{5.22}{.04}   & \ablval{5.21}{.05}   & \ablval{5.12}{.02}   & \ablval{5.26}{.04}  \\
		\midrule
		\multirow{2}{*}{Pred.-Prey Medium} & w/  & \ablval{27.2}{0.6}   & \ablbest{23.3}{1.5}  & \ablval{17.9}{2.5}   & \ablbest{13.1}{0.23} & \ablbest{12.8}{0.1}  & \ablbest{12.6}{0.2}  & \ablbest{12.6}{0.2} \\
		                                   & w/o & \ablbest{26.8}{0.34} & \ablval{27.2}{2.9}   & \ablbest{15.8}{4.5}  & \ablval{14.3}{1.5}   & \ablval{13.3}{0.4}   & \ablval{13.1}{0.2}   & \ablval{13.3}{0.4}  \\
		\midrule
		\multirow{2}{*}{SHAPES \(\times 10^{2}\)} & w/  & \ablbest{-5.3}{.20}  & \ablbest{-4.8}{.82}  & \ablbest{-4.7}{.30}  & \ablbest{-3.9}{.64}  & \ablbest{-3.1}{.18}  & \ablbest{-3.0}{.32}  & \ablbest{-4.6}{.22} \\
		                                   & w/o & \ablval{-7.0}{.40}   & \ablval{-7.2}{.55}   & \ablval{-6.8}{.79}   & \ablval{-5.9}{.76}   & \ablval{-6.4}{.44}   & \ablval{-7.0}{.43}   & \ablval{-7.2}{.44}  \\
		\bottomrule
	\end{tabular*}
\end{table}

\paragraph{Navigation} In order to illustrate that the method could scale to more agents, we also illustrate SLIM using \href{https://imgur.com/a/UH4kBVn}{4}, \href{https://imgur.com/a/fOPwhWa}{20}, \href{https://imgur.com/a/4SqKSwI}{24}, \href{https://imgur.com/a/rsAkebg}{32} and \href{https://imgur.com/a/QYYeXES}{42} agents (anonymised links). These illustrations show agents successfully reaching their individual goals while avoiding collisions despite not seeing each other.

Due to computational constraints, we have only benchmarked SLIM against TarMAC (the strongest baseline when using 4 agents) in the 20 agents setting. In this setting, TarMAC gets a reward of \(0.02 \pm 0.01\) while SLIM gets \(0.95 \pm 0.003\).

\section*{Reproducibility}

We publicly released the code to reproduce all our experiments on \href{https://github.com/alexicanesse/Decoupling-Communication-from-Policy-Robust-MARL-under-Bandwidth-Constraints}{GitHub}. All experiments can be run using a simple script. Environment files are available with all the libraries version used to be able to exactly reproduce our setting. 

\section*{Licenses}\label{sec:licenses}

We release our implementation under the \href{https://choosealicense.com/licenses/mit/}{MIT licence}. We do not redistribute the code of the baselines. CommFormer is available through the \href{https://choosealicense.com/licenses/apache-2.0/}{Apache License 2.0}, IC3Net and TarMAC through the \href{https://choosealicense.com/licenses/mit/}{MIT License}, and CommNet through the BSD license.

The Predator-Prey and Traffic Junction environments are available using the \href{https://choosealicense.com/licenses/mit/}{MIT License}, VMAS through the \href{https://choosealicense.com/licenses/gpl-3.0/}{GPL3 license} and we reimplemented the SHAPES environment based on the description provided in the original paper, thus we release it under the \href{https://choosealicense.com/licenses/mit/}{MIT License}.

\section*{Environments Details}

\paragraph*{Predator-Prey.} The Predator-Prey environment is a multi-agent grid world where agents (predators) must cooperate to locate a prey with limited vision (local observability) centred on them. Each agent gets a negative reward until it reaches the prey. Agents can move in four directions (up, down, left, right) or stay in place. They benefit from a strong communication since knowing where the prey is not gives information about its location. Since the prey cannot move, the environment is not jointly fully observable: information about previously seen locations is information on the state, that could be missing from the current joint observation. To address this, we enable the message history cache for this environment and provide an ablation study in~\Cref{sec:ablation_cache} to assess its impact.

\paragraph*{Traffic Junction.} The Traffic Junction environment simulates intersections that agents \textit{without any vision} need to cross. Agents enter the environment stochastically and navigate along randomly assigned routes. The lack of vision makes communication essential to avoid collisions, making this environment a suitable benchmark for communication MARL. In this environment, agents do not have to navigate: they can either go forward along their route or wait. The reward strongly incentivise collision avoidance while lightly encouraging efficiency: agents receive an increasingly negative reward for each time step they are active in the environment, and a large negative reward if they collide with another car. Since the concatenation of all agents' local observations (cars position) fully characterises the global configuration of the system, the environment is categorised as a decentralised Markov decision process (Dec-MDP). Consequently, the message history cache is deactivated for this environment.

\paragraph*{Navigation} The navigation environment~\cite{bettini2022vmas} implements a bounded continuous world where \(n\) randomly spawned agents must navigate to individual randomly spawned goals without colliding with other agents, without seeing each other. Each agent only sees its position, speed and relative distance to its goal. Actions are acceleration in different directions and the environment simulates momentum, making fine control complicated. Agents receive a reward proportional to the distance gained to their goal at each time step, and a large negative reward if they collide with another agent.

\paragraph*{SHAPES.} In order to verify the effectiveness of the message history cache in~\Cref{sec:ablation_cache}, we introduced a variation of the simple SHAPES environment~\cite{andreas2016neural, das2019tarmac}. Images containing coloured shapes are generated. Agents spawn at random positions on a random image and must find a (possibly different) random colour each whilst only being able to observe their surroundings. This environment is not partially jointly fully observable. Since agents have different goals, information on what other agents have seen in the past can contain information about another agent's goal. Each agent receives a reward of \(10^{-2}\) until they reach their target, at which point they receive a reward of 0.\label{app:environments}



\end{document}